\documentclass[apj,numberedappendix]{emulateapj}
\usepackage{amssymb,amsmath}
\usepackage{graphicx}
\usepackage{appendix}
\usepackage{natbib}
\usepackage{hyperref}
\usepackage{enumerate}
\usepackage{multirow}
\usepackage{bm}

\newcommand{\etal}{et~al.~}

\altaffiltext{\hubble}{Hubble Fellow}
\altaffiltext{\MIT}{Kavli Institute for Astrophysics and Space Research, Massachusetts Institute of Technology, 77 Massachusetts Avenue, Cambridge, MA 02139}
\altaffiltext{\FNAL}{Fermi National Accelerator Laboratory, Batavia, IL 60510-0500}
\altaffiltext{\KICPChicago}{Kavli Institute for Cosmological Physics, University of Chicago, 5640 South Ellis Avenue, Chicago, IL 60637}
\altaffiltext{\AAUChicago}{Department of Astronomy and Astrophysics, University of Chicago, 5640 South Ellis Avenue, Chicago, IL 60637}
\altaffiltext{\CfA}{Harvard-Smithsonian Center for Astrophysics, 60 Garden Street, Cambridge, MA 02138}
\altaffiltext{\UChicago}{University of Chicago, 5640 South Ellis Avenue, Chicago, IL 60637}
\altaffiltext{\KIPAC}{Kavli Institute for Particle Astrophysics and Cosmology, Stanford University, 452 Lomita Mall, Stanford, CA 94305}
\altaffiltext{\Stanford}{Department of Physics, Stanford University, 382 Via Pueblo Mall, Stanford, CA 94305}
\altaffiltext{\SLAC}{SLAC National Accelerator Laboratory, 2575 Sand Hill Road, Menlo Park, CA 94025}
\altaffiltext{\Harvard}{Department of Physics, Harvard University, 17 Oxford Street, Cambridge, MA 02138}
\altaffiltext{\PhysicsUChicago}{Department of Physics, University of Chicago, 5640 South Ellis Avenue, Chicago, IL 60637}
\altaffiltext{\ANL}{Argonne National Laboratory, 9700 S. Cass Avenue, Argonne, IL, USA 60439}
\altaffiltext{\Munich}{Department of Physics, Ludwig-Maximilians-Universit\"{a}t, Scheinerstr.\ 1, 81679 M\"{u}nchen, Germany}
\altaffiltext{\ExcellenceCluster}{Excellence Cluster Universe, Boltzmannstr.\ 2, 85748 Garching, Germany}
\altaffiltext{\Miss}{Department of Physics and Astronomy, University of Missouri, 5110 Rockhill Road, Kansas City, MO 64110}
\altaffiltext{\EFIChicago}{Enrico Fermi Institute, University of Chicago, 5640 South Ellis Avenue, Chicago, IL 60637}
\altaffiltext{\NIST}{NIST Quantum Devices Group, 325 Broadway Mailcode 817.03, Boulder, CO, USA 80305}
\altaffiltext{\PUC}{Departamento de Astronomia y Astrosifica, Pontificia Universidad Catolica, Chile}
\altaffiltext{\Caltech}{California Institute of Technology, 1200 E. California Blvd., Pasadena, CA 91125}
\altaffiltext{\McGill}{Department of Physics, McGill University, 3600 Rue University, Montreal, Quebec H3A 2T8, Canada}
\altaffiltext{\illast}{Astronomy Department, University of Illinois at Urbana-Champaign, 1002 W.\ Green Street, Urbana, IL 61801 USA}
\altaffiltext{\illphy}{Department of Physics, University of Illinois Urbana-Champaign, 1110 W.\ Green Street, Urbana, IL 61801 USA}
\altaffiltext{\Berkeley}{Department of Physics, University of California, Berkeley, CA 94720}
\altaffiltext{\UFlorida}{Department of Astronomy, University of Florida, Gainesville, FL 32611}
\altaffiltext{\Colorado}{Department of Astrophysical and Planetary Sciences and Department of Physics, University of Colorado, Boulder, CO 80309}
\altaffiltext{\Davis}{Department of Physics, University of California, One Shields Avenue, Davis, CA 95616}
\altaffiltext{\LBNL}{Physics Division, Lawrence Berkeley National Laboratory, Berkeley, CA 94720}
\altaffiltext{\Arizona}{Steward Observatory, University of Arizona, 933 North Cherry Avenue, Tucson, AZ 85721}
\altaffiltext{\Michigan}{Department of Physics, University of Michigan, 450 Church Street, Ann Arbor, MI, 48109}
\altaffiltext{\MPE}{Max-Planck-Institut f\"{u}r extraterrestrische Physik, Giessenbachstr.\ 85748 Garching, Germany}
\altaffiltext{\Minnesota}{Physics Department, University of Minnesota, 116 Church Street S.E., Minneapolis, MN 55455}
\altaffiltext{\STScI}{Space Telescope Science Institute, 3700 San Martin Dr., Baltimore, MD 21218}
\altaffiltext{\CaseWestern}{Physics Department, Center for Education and Research in Cosmology and Astrophysics, Case Western Reserve University, Cleveland, OH 44106}
\altaffiltext{\SAIC}{Liberal Arts Department, School of the Art Institute of Chicago, 112 S Michigan Ave, Chicago, IL 60603}
\altaffiltext{\LLNL}{Institute of Geophysics and Planetary Physics, Lawrence Livermore National Laboratory, Livermore, CA 94551}
\altaffiltext{\Dunlap}{Dunlap Institute for Astronomy \& Astrophysics, University of Toronto, 50 St George St, Toronto, ON, M5S 3H4, Canada}
\altaffiltext{\Toronto}{Department of Astronomy \& Astrophysics, University of Toronto, 50 St George St, Toronto, ON, M5S 3H4, Canada}
\altaffiltext{\BCCP}{Berkeley Center for Cosmological Physics, Department of Physics, University of California, and Lawrence Berkeley National Labs, Berkeley, CA 94720}
\altaffiltext{\CTIO}{Cerro Tololo Inter-American Observatory, La Serena, Chile}

\def\MIT{1}
\def\FNAL{2}
\def\KICPChicago{3}
\def\AAUChicago{4}
\def\CfA{5}
\def\UChicago{6}
\def\KIPAC{7}
\def\Stanford{8}
\def\SLAC{9}
\def\Harvard{10}
\def\ANL{11}
\def\Munich{12}
\def\ExcellenceCluster{13}
\def\Miss{14}
\def\PhysicsUChicago{15}
\def\EFIChicago{16}
\def\NIST{17}
\def\PUC{18}
\def\Caltech{19}
\def\McGill{20}
\def\illast{21}
\def\illphy{22}
\def\Berkeley{23}
\def\UFlorida{24}
\def\Colorado{25}
\def\Davis{26}
\def\LBNL{27}
\def\Arizona{28}
\def\Michigan{29}
\def\MPE{30}
\def\Minnesota{31}
\def\STScI{32}
\def\CaseWestern{33}
\def\SAIC{34}
\def\LLNL{35}
\def\Dunlap{36}
\def\Toronto{37}
\def\BCCP{38}
\def\CTIO{39}

\def\hubble{$\dagger$}

\begin{document}


\title{The Redshift Evolution of the Mean Temperature, Pressure, \\and Entropy Profiles in 80 SPT-Selected Galaxy Clusters}



\author{
M.~McDonald\altaffilmark{\MIT,\hubble}
B.~A.~Benson\altaffilmark{\FNAL,\KICPChicago,\AAUChicago},
A. Vikhlinin\altaffilmark{\CfA},
K.~A.~Aird\altaffilmark{\UChicago},
S.~W.~Allen\altaffilmark{\KIPAC,\Stanford,\SLAC},
M.~Bautz\altaffilmark{\MIT},
M.~Bayliss\altaffilmark{\Harvard,\CfA},
L.~E.~Bleem\altaffilmark{\KICPChicago,\ANL},
S.~Bocquet\altaffilmark{\Munich,\ExcellenceCluster},
M.~Brodwin\altaffilmark{\Miss},
J.~E.~Carlstrom\altaffilmark{\KICPChicago,\EFIChicago,\PhysicsUChicago,\ANL,\AAUChicago}, 
C.~L.~Chang\altaffilmark{\KICPChicago,\AAUChicago,\ANL}, 
H.~M. Cho\altaffilmark{\NIST}, 
A.~Clocchiatti\altaffilmark{\PUC},
T.~M.~Crawford\altaffilmark{\KICPChicago,\AAUChicago},
A.~T.~Crites\altaffilmark{\KICPChicago,\AAUChicago,\Caltech},
T.~de~Haan\altaffilmark{\McGill},
M.~A.~Dobbs\altaffilmark{\McGill},
R.~J.~Foley\altaffilmark{\illast,\illphy},
W.~R.~Forman\altaffilmark{\CfA},
E.~M.~George\altaffilmark{\Berkeley},
M.~D.~Gladders\altaffilmark{\KICPChicago,\AAUChicago},
A.~H.~Gonzalez\altaffilmark{\UFlorida},
N.~W.~Halverson\altaffilmark{\Colorado},
J.~Hlavacek-Larrondo\altaffilmark{\KIPAC,\Stanford},
G.~P.~Holder\altaffilmark{\McGill},
W.~L.~Holzapfel\altaffilmark{\Berkeley},
J.~D.~Hrubes\altaffilmark{\UChicago},
C.~Jones\altaffilmark{\CfA},
R.~Keisler\altaffilmark{\KICPChicago,\PhysicsUChicago},
L.~Knox\altaffilmark{\Davis},
A.~T.~Lee\altaffilmark{\Berkeley,\LBNL},
E.~M.~Leitch\altaffilmark{\KICPChicago,\AAUChicago},
J.~Liu\altaffilmark{\Munich,\ExcellenceCluster},
M.~Lueker\altaffilmark{\Berkeley,\Caltech},
D.~Luong-Van\altaffilmark{\UChicago},
A.~Mantz\altaffilmark{\KICPChicago},
D.~P.~Marrone\altaffilmark{\Arizona},
J.~J.~McMahon\altaffilmark{\Michigan},
S.~S.~Meyer\altaffilmark{\KICPChicago,\EFIChicago,\PhysicsUChicago,\AAUChicago},
E.~D.~Miller\altaffilmark{\MIT},
L.~Mocanu\altaffilmark{\KICPChicago,\AAUChicago},
J.~J.~Mohr\altaffilmark{\Munich,\ExcellenceCluster,\MPE},
S.~S.~Murray\altaffilmark{\CfA},
S.~Padin\altaffilmark{\KICPChicago,\AAUChicago,\Caltech},
C.~Pryke\altaffilmark{\Minnesota}, 
C.~L.~Reichardt\altaffilmark{\Berkeley},
A.~Rest\altaffilmark{\STScI},
J.~E.~Ruhl\altaffilmark{\CaseWestern}, 
B.~R.~Saliwanchik\altaffilmark{\CaseWestern}, 
A.~Saro\altaffilmark{\Munich},
J.~T.~Sayre\altaffilmark{\CaseWestern}, 
K.~K.~Schaffer\altaffilmark{\KICPChicago,\EFIChicago,\SAIC}, 
E.~Shirokoff\altaffilmark{\Berkeley,\Caltech}, 
H.~G.~Spieler\altaffilmark{\LBNL},
B.~Stalder\altaffilmark{\CfA},
S.~A.~Stanford\altaffilmark{\Davis,\LLNL},
Z.~Staniszewski\altaffilmark{\CaseWestern,\Caltech},
A.~A.~Stark\altaffilmark{\CfA}, 
K.~T.~Story\altaffilmark{\KICPChicago,\PhysicsUChicago},
C.~W.~Stubbs\altaffilmark{\Harvard,\CfA}, 
K.~Vanderlinde\altaffilmark{\Dunlap,\Toronto},
J.~D.~Vieira\altaffilmark{\illast,\illphy,\Caltech},
R.~Williamson\altaffilmark{\KICPChicago,\AAUChicago,\Caltech}, 
O.~Zahn\altaffilmark{\Berkeley,\BCCP},
A.~Zenteno\altaffilmark{\CTIO}
}

\email{Email: mcdonald@space.mit.edu}   


\begin{abstract}
We present the results of an X-ray analysis of 80 galaxy clusters selected in the 2500 deg$^2$ South Pole Telescope survey and observed with the \emph{Chandra X-ray Observatory}. We divide the full sample into subsamples of $\sim$20 clusters based on redshift and central density, performing a joint X-ray spectral fit to all clusters in a subsample simultaneously, assuming self-similarity of the temperature profile. This approach allows us to constrain the shape of the temperature profile over $0 < r < 1.5R_{500}$, which would be impossible on a per-cluster basis, since the observations of individual clusters have, on average, 2000 X-ray counts. The results presented here represent the first constraints on the evolution of the average temperature profile from $z = 0$ to $z = 1.2$.
We find that high-$z$ ($0.6 < z < 1.2$) clusters are slightly ($\sim$30\%) cooler both in the inner ($r<0.1R_{500}$) and outer ($r>R_{500}$) regions than their low-$z$ ($0.3 < z<0.6$) counterparts. Combining the average temperature profile with measured gas density profiles from our earlier work, we infer the average pressure and entropy profiles for each subsample. 
Confirming earlier results from this data set, we find an absence of strong cool cores at high $z$, manifested in this analysis as a significantly lower observed pressure in the central $0.1R_{500}$ of the high-$z$ cool-core subset of clusters compared to the low-$z$ cool-core subset.
Overall, our observed pressure profiles agree well with earlier lower-redshift measurements, suggesting minimal redshift evolution in the pressure profile outside of the core. We find no measurable redshift evolution in the entropy profile at $r\lesssim0.7R_{500}$ -- this may reflect a long-standing balance between cooling and feedback over long timescales and large physical scales. We observe a slight flattening of the entropy profile at $r\gtrsim R_{500}$ in our high-$z$ subsample. This flattening is consistent with a temperature bias due to the enhanced ($\sim$3$\times$) rate at which group-mass ($\sim$2\,keV) halos, which would go undetected at our survey depth, are accreting onto the cluster at $z\sim1$. This work demonstrates a powerful method for inferring spatially-resolved cluster properties in the case where individual cluster signal-to-noise is low, but the number of observed clusters is high.


\end{abstract}

\keywords{galaxies: clusters: general -- galaxies: clusters: intracluster medium -- cosmology: early universe --  X-rays: galaxies: clusters \vspace{-0.2in}}

\section{Introduction}
\setcounter{footnote}{0}

Galaxy clusters, despite what the name implies, consist primarily of matter that is not in galaxies. A typical cluster is well-modeled by a central dark matter halo ($\sim$85\% by mass) and a diffuse, optically-thin plasma ($\sim$15\% by mass). The response of this hot ($\gtrsim10^7$~K) plasma, known as the intracluster medium (ICM), to the evolving gravitational potential is one of our best probes of the current state and evolution of galaxy clusters. 
X-ray imaging and spectroscopy of the ICM allow estimates of the cluster mass profile via the spectroscopic temperature and gas density \citep[e.g.,][]{forman82,nevalainen00,sanderson03,arnaud05,kravtsov06, vikhlinin06a, arnaud07}, the enrichment history of the cluster via the ICM metallicity \citep[e.g.,][]{deyoung78,matteucci86,deplaa07,bregman10,bulbul12}, the cooling history via the cooling time or entropy \citep[e.g.,][]{white97, peres98, cavagnolo08, mcdonald13b}, the feedback history via the presence of X-ray bubbles \citep[e.g.,][]{rafferty06, mcnamara07,rafferty08,hlavacek12}, and the current dynamical state and merger history of the cluster via the X-ray morphology \citep[e.g.,][]{jones79,mohr95,roettiger96,schuecker01,jeltema05,nurgaliev13}.

While there is much diversity in the ICM from cluster to cluster, it is valuable to determine if there are broad similarities in clusters of a given mass and redshift. The construction of a ``Universal'' pressure profile, for example, can allow comparisons to simulated galaxy clusters, as well as provide a functional form for matched-filtering algorithms, such as those that are used to select galaxy clusters using the Sunyaev-Zel'dovich  \citep[SZ;][]{sunyaev72} effect. Much effort has been made to quantify the average temperature \citep[e.g.,][]{loken02, vikhlinin06a, pratt07, leccardi08a, baldi12}, entropy \citep[e.g.,][]{voit05, piffaretti05,cavagnolo09, pratt10}, and pressure \citep[e.g.][]{arnaud10, sun11,bonamente12,planck13} profiles for low-redshift galaxy groups and clusters based on both X-ray and SZ selection. In all cases, the average profiles have a substantial amount of scatter at $r\lesssim0.2R_{500}$, due to the presence (or lack) of a cool, dense core \citep[e.g.,][]{vikhlinin06a,cavagnolo09,arnaud10}, but collapse onto the self-similar expectation at larger radii. This suggests that non-gravitational processes (e.g., cooling, AGN feedback) are important in the central region of the cluster while gravity is the dominant force in the outer region.

While the aforementioned studies have made significant progress in quantifying the average temperature, entropy, and pressure profiles of galaxy groups and clusters, they have focused almost entirely on low-redshift ($z\lesssim0.2$) systems. This is in part due to the relative ease with which one can measure the temperature profile in nearby systems, but also due to the fact that, until recently, large, well-selected samples of galaxy clusters at high redshift did not exist. This has changed, with the recent success of large SZ surveys from the Atacama Cosmology Telescope \citep[ACT;][]{act11,act13}, \emph{Planck} \citep{planck11,planck13b}, and the South Pole Telescope \citep[SPT;][]{vanderlinde10,reichardt13}. These surveys have discovered hundreds of new galaxy clusters at $z>0.3$, allowing the study of galaxy cluster evolution for the first time out to $z>1$ using large, homogeneous data sets.

In this paper, we present a joint-fit spectroscopic analysis of 80 SPT-selected galaxy clusters in the SPT-XVP sample \citep[][Benson \etal in prep]{mcdonald13b}. Utilizing uniform-depth X-ray observations of these clusters we can, for the first time, constrain the redshift evolution of the average ICM temperature, entropy, and pressure profiles. We present the details of this analysis in \S2, including the resulting projected and deprojected temperature profiles in \S2.2 and \S2.3, respectively. In \S3 we infer the average pressure (\S3.1) and entropy (\S3.2) profiles. In \S4 we discuss the implications of the observed evolution, specifically in the inner $\sim$100\,kpc and outskirts ($r\gtrsim R_{500}$) of the mean pressure and entropy profiles, and assess any potential biases in our analysis. Finally, we summarize these results in \S5 before suggesting future applications of these data. Throughout this work, we assume H$_0$=70 km s$^{-1}$ Mpc$^{-1}$, $\Omega_M$ = 0.27, and $\Omega_{\Lambda}$ = 0.73.


\section{Data and Analysis}

\subsection{Sample}
\begin{figure}[htb]
\centering
\includegraphics[width=0.49\textwidth,trim=1.4cm 0cm 0.1cm 0cm,clip=true]{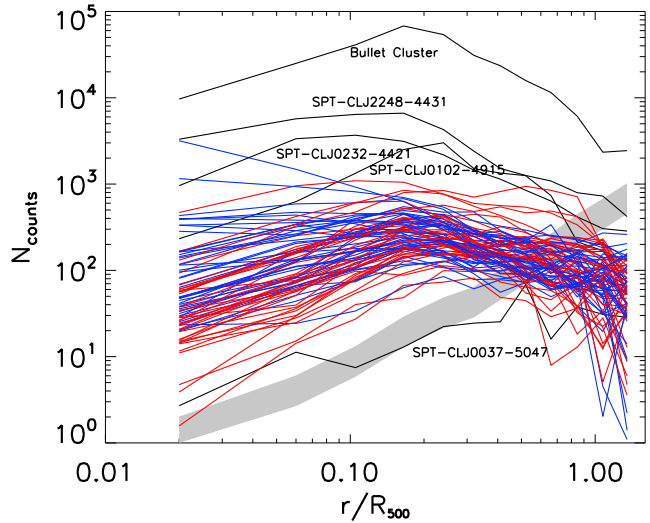} 
\caption{Number of X-ray counts per annulus in the energy range 0.7--7.0 keV for 85 SPT-selected galaxy clusters observed with \emph{Chandra}. Blue and red curves correspond to clusters classified as cool core and non-cool core, respectively (see \S2.1). The shaded gray band represents the measured Galactic and extragalactic background in each annulus. Five clusters which are either too high or too low signal-to-noise to be used in the stacking analysis are shown in black. Overall, the remaining 80 clusters have a tight distribution of counts at $r>0.2R_{500}$. We stress that this has not been scaled for exposure time -- we have nearly uniform depth ($\sim$2000 counts per cluster) over this full sample.
}
\label{fig:counts}
\end{figure}

The majority of the observations for this program were obtained as part of a \emph{Chandra X-ray Visionary Project} to observe the 80 most massive SPT-selected galaxy clusters at $0.4 < z < 1.2$ (PI: B.\ Benson). This survey is described in more detail in \cite{mcdonald13b} and Benson \etal (in prep). We begin our sample definition by identifying 91 galaxy clusters that are detected in the SPT 2500 deg$^2$ survey and have been observed with \emph{Chandra}. We first exclude four clusters which are detected with the SPT at $S/N < 6$ (SPT-CLJ0236-4938, SPT-CLJ0522-5818, SPT-CLJ2011-5725, SPT-CLJ2332-5053), which gives our sample a uniform SZ selection at $S/N \sim 6.5$.
We exclude two additional clusters (SPT-CLJ0330-5228, SPT-CLJ0551-5709) which suffer from severe projection effects due to extended foreground sources (i.e., nearby, low-mass groups).
In the remaining 85 systems, we identify a tight distribution of X-ray photon counts at $r>0.2R_{500}$ (see Figure \ref{fig:counts}), and exclude four clusters with exceptionally high signal-to-noise (SPT-CLJ0658-5556 (the Bullet Cluster), SPT-CLJ2248-4431, SPT-CLJ0232-4421, SPT-CLJ0102-4915), which could dominate the stacking analysis, and one cluster with very low counts (SPT-CLJ0037-5047), which will not contribute meaningful signal. What remains is a sample of 80 clusters that occupy a tight sequence in signal-to-noise at large radii. These 80 clusters define our sample of massive (M$_{500} \gtrsim 3\times10^{14}$ E($z$)$^{-1}$ M$_{\odot}$) galaxy clusters with uniform-depth X-ray imaging, spanning the redshift range $0.3 \lesssim z \lesssim 1.2$. We note that, in the outermost annulus, the combined Galactic and extragalactic background is roughly an order of magnitude brighter than the source emission. However by combining $\sim$20 spectra ($\sqrt{20}$ improvement) and joint-fitting the background in the on- and off-source regions ($\sqrt{2}$ improvement), we improve the signal-to-noise from $\sim$3 per spectrum to $\sim$20 -- enough to constrain the spectroscopic temperature to within $\sim$10\%. 

This sample of 80 clusters was divided into six subsamples, based on individual cluster redshift and the presence (or lack) of a cool core, in order to probe the redshift evolution and cooling-dependence of the universal temperature, pressure, and entropy profiles. For simplicity, and so that subsamples are of equivalent signal-to-noise, we divide the low-redshift and high-redshift subsamples in half, with the 50\% ``cuspiest'' clusters, where cuspiness ($\alpha$) is defined as the slope of the gas density profile at 0.04R$_{500}$ \citep{vikhlinin07}, making up the cool core (CC) subsample and the 50\% least cuspy clusters defining the non-cool core (NCC) subsample. This yields the six subsamples summarized in Table \ref{table:subsamples}.
The choice of $z=0.6$ as a dividing line was motivated by the desire to have an equal number of clusters in both the high-$z$ and low-$z$ bins. The mean redshift for the two redshift bins are $\left<z\right>_{low} =  0.46$ and $\left<z\right>_{high} =  0.82$, which provides a broad baseline for comparison to previous studies at $\left<z\right> \sim 0.1$ (see \S1).

\begin{deluxetable}{c c c c c c}[htb]
\tablecaption{Cluster Subsamples}
\tablehead{
\colhead{Subsample} &
\colhead{$z$ range} &
\colhead{$\alpha$ range} & 
\colhead{N$_{cluster}$} & 
\colhead{$\left<z\right>$} &
\colhead{$\left<M_{500}\right>^*$}
}
\startdata
low-$z$ & $z<0.6$ & -- & 40 & 0.46 & $5.5 \pm 0.3$\\
low-$z$, CC & $z<0.6$ & $\alpha > 0.39$ & 19 & 0.48 & $5.3 \pm 0.5$\\
low-$z$, NCC & $z<0.6$ & $\alpha < 0.39$ & 21 & 0.45 & $5.7 \pm 0.4$\\ 
\\
high-$z$ & $z>0.6$ & -- & 40 & 0.82 & $4.2 \pm 0.2$\\
high-$z$, CC & $z>0.6$ & $\alpha > 0.39$ & 20 & 0.80 & $3.9 \pm 0.3$\\
high-$z$, NCC & $z>0.6$ & $\alpha < 0.39$ & 20 & 0.84 & $4.4 \pm 0.3$\\
\enddata
\tablecomments{$^*$: in units of 10$^{14}$ M$_{\odot}$. Error on the mean is shown.}
\label{table:subsamples}
\end{deluxetable}

\begin{figure*}[htb]
\centering
\includegraphics[width=0.9\textwidth,trim=1.5cm 12cm 1cm 0cm,clip=true]{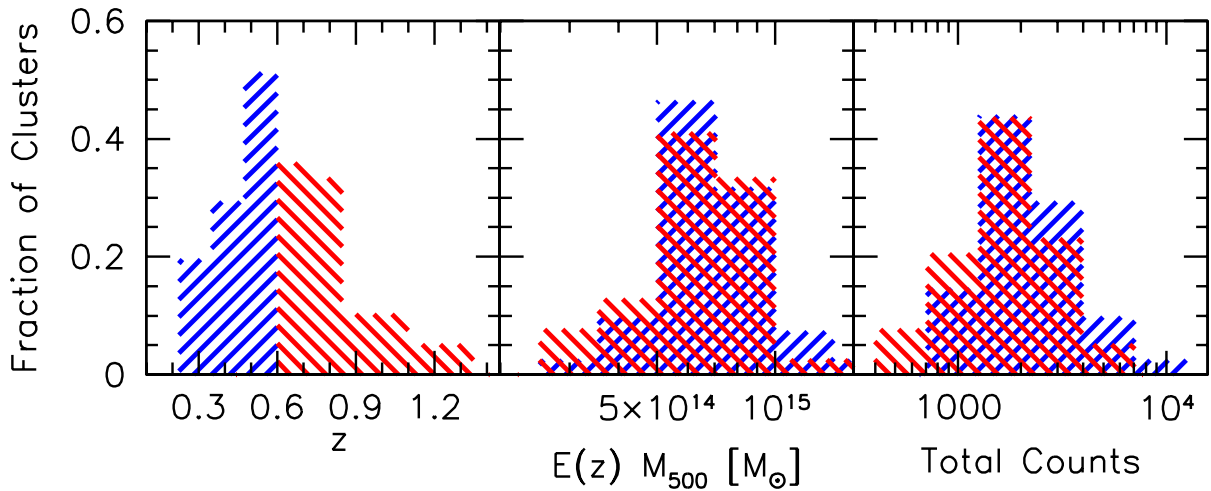} 
\caption{Distribution of redshift ($z$), mass (M$_{500}$) and total X-ray counts for the 80 clusters used in this work. Here M$_{500}$ is computed assuming the Y$_X$--M relation from \cite{vikhlinin09a}, as described in \cite{mcdonald13b} and Benson \etal (in prep). We show separately the distributions for low-$z$ (blue; $z<0.6$) and high-$z$ (red; $z>0.6$) clusters, demonstrating the similarity in mass and data quality between these two subsamples.}
\label{fig:sample}
\end{figure*}

Global cluster properties (e.g., M$_{500}$, kT$_{500}$) used in this work are derived in Benson \etal (in prep),
following closely the procedures described in \cite{andersson11}. Briefly, we estimate R$_{500}$, the radius within which the average enclosed density is 500 times the critical density, by iteratively adjusting R$_{500}$ until the measured Y$_X$ ($\equiv M_g \times kT$) satisfies the Y$_{X,500}-$M$_{500}$ relation \citep{vikhlinin09a}, which assumes a purely self-similar evolution, M$_{500}$~$\propto$~Y$_{X,500}$~$E(z)^{-2/5}$. Once the radius converges, kT$_{500}$ is measured in a core-excised annulus from 0.15R$_{500}$ $<$ $r$ $<$ R$_{500}$.  Further details on the derivation of global X-ray properties can be found in \cite{andersson11}. We point out that we have also examined the effects of using an M$_{gas}$-derived temperature to normalize the temperature profiles, following \cite{vikhlinin06a}, and confirm that the results we will present below are independent of our choice of normalization.

In Figure \ref{fig:sample} we show the distributions of redshift ($z$), mass (M$_{500}$), and total X-ray counts for the full sample of 80 clusters, as well as the low- and high-redshift subsamples. Here M$_{500}$ is computed assuming the Y$_X$--M relation from \cite{vikhlinin09a}, as described in \cite{mcdonald13b} and Benson \etal (in prep). This figure demonstrates that the low-$z$ and high-$z$ subsamples are comprised of similar clusters, in terms of their total mass, and have been observed to similar depths, allowing a fair comparison.

\subsection{Joint Spectral Fitting}

For each cluster, we extract spectra in 12 annuli spanning $0 < r < 1.5R_{500}$. At $r>0.3R_{500}$, we use logarithmically-spaced bins, while interior to $0.3R_{500}$, we choose a binning scheme which achieves fine sampling while also maintaining suitable signal-to-noise per annulus. The bin edges are tabulated in Table \ref{table:data}.
This two-part binning scheme yields roughly equal signal-to-noise in all bins, without requiring overly narrow/wide bins at any radius. The total number of annuli was chosen, through trial and error, so that the cluster-to-cluster scatter in $kT/kT_{500}$ within a given annulus is similar to the fit uncertainty (i.e., there is negligible improvement in scatter from widening the bins). The number of counts per bin, for each cluster, is shown in Figure \ref{fig:counts}. This figure demonstrates that, even in our outermost bin ($1.2R_{500} < r < 1.5R_{500}$), we have $\sim$100 X-ray counts per cluster spectrum.

\begin{figure*}[htb]
\centering
\includegraphics[width=0.99\textwidth]{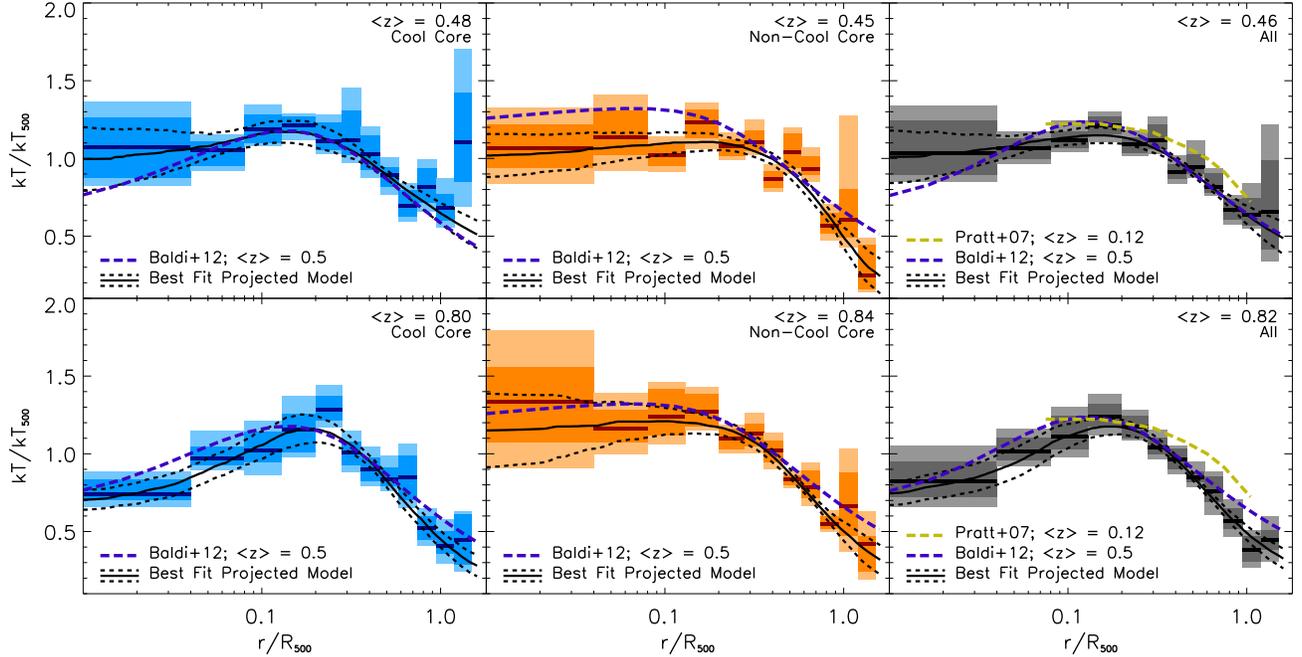}
\caption{Average projected temperature profiles for 80 SPT-selected clusters. In all panels, the colored regions represent the data, with dark and light regions corresponding to 1$\sigma$ and 90\% confidence, respectively, while the solid line corresponds to the median value. These projected temperature profiles are derived from joint spectral fits, as described in \S2.2. The best fit projected profile is shown in solid black, with the 1$\sigma$ uncertainty range in dotted black (see \S2.3). For comparison, average projected temperature profiles from previous studies \citep{pratt07,baldi12} are shown where applicable. In general, the agreement with previous works at similar redshifts is qualitatively good.}
\label{fig:univT}
\end{figure*}

For each cluster, we extract spectra in each of these annuli, using \textsc{ciao} v4.6 and \textsc{caldb} v4.6.1.1, along with accompanying background (both blank-sky and adjacent to source) and response files. Blank-sky background spectra are rescaled based on the observed 9.5--12\,keV flux and subtracted from both on- and off-source spectra, which are then simultaneously modeled in the procedures outlined below. All spectra are binned in energy with wide bins at low ($\lesssim$1\,keV) and high ($\gtrsim$4\,keV) energy where the signal-to-noise is low. The average binning was $\Delta(\log_{10}E) = 0.1$. 
Given a subsample of $N$ clusters (e.g., 40 low-$z$ clusters), we randomly draw $N$ clusters, allowing repeats (i.e., bootstrap analysis). This means that, in some cluster realizations, the contributions from cluster $i$ may be double-counted, while cluster $j$ will be excluded. The spectra for these $N$ clusters are simultaneously fit over the energy range 0.5--10.0~keV with \textsc{xspec} \citep[v12.8.0;][]{arnaud96}, using a combination of a single-temperature plasma \citep[\textsc{apec};][]{smith01}, a soft X-ray background contribution (\textsc{apec}, $kT=0.18$~keV), a hard X-ray background contribution (\textsc{bremss}, $kT=40$~keV), and a Galactic absorption model (\textsc{phabs})\footnote{\scriptsize http://heasarc.gsfc.nasa.gov/docs/xanadu/xspec/manual/ \\ XspecModels.html}. These additional soft (Galactic foreground) and hard (unresolved CXB) emission models account for any residual emission after the blank-sky backgrounds are subtracted. The various free parameters of the plasma model are constrained as described in Table \ref{table:xspec}.

\begin{deluxetable}{c c c c}[htb]
\tablecaption{Parameters for Joint Spectral Fitting}
\tablehead{
\colhead{Param.} &
\colhead{Description} &
\colhead{Spectrum 0} & 
\colhead{Spectrum $i$}
}
\startdata
N$_H$ & H column density & constrained$^1$ & constrained$^1$ \\
$kT$ & plasma temperature & free ($kT_0$) & $kT_0\left(\frac{kT_{500,i}}{kT_{500,0}}\right)$ \\
z & redshift & constrained$^2$ & constrained$^2$ \\
Z & metal abundance & free$^3$ & free$^3$ \\
N & normalization$^4$ & free$^{ ~}$ & free$^{ ~}$ 
\enddata
\tablecomments{$^1$:\,Constrained to within 15\% of the Galactic value from \cite{kalberla05}. Other spectra of the same field share the same value of N$_H$. $^2$:\,Constrained to within 5\% of the optical redshift \citep[see ][]{song12, ruel13}. Other spectra of the same cluster share the same value of $z$. $^3$:\,Limited range to $0 < Z < 2Z_{\odot}$. $^4$: $\frac{10^{-14}}{4\pi(D_A(1+z))^2}\int n_en_HdV$}
\label{table:xspec}
\end{deluxetable}

\begin{deluxetable*}{c c c c c c c}[htb]
\tablecaption{Average Projected Temperature Profiles}
\tablehead{
\colhead{} &
\colhead{low-$z$} &
\colhead{low-$z$, CC} & 
\colhead{low-$z$, NCC} & 
\colhead{high-$z$} &
\colhead{high-$z$, CC} & 
\colhead{high-$z$, NCC}\\
\colhead{$r$/R$_{500}$} & 
\colhead{$kT/kT_{500}$} & 
\colhead{$kT/kT_{500}$} & 
\colhead{$kT/kT_{500}$} & 
\colhead{$kT/kT_{500}$} & 
\colhead{$kT/kT_{500}$} & 
\colhead{$kT/kT_{500}$} 
}
\startdata
0.00--0.04 & 1.03$_{-0.14}^{+0.21}$ & 1.07$_{-0.19}^{+0.19}$ & 1.06$_{-0.12}^{+0.16}$ & 0.79$_{-0.07}^{+0.11}$ & 0.74$_{-0.04}^{+0.09}$ & 1.32$_{-0.35}^{+0.35}$ \\
0.04--0.08 & 1.07$_{-0.07}^{+0.06}$ & 1.05$_{-0.10}^{+0.06}$ & 1.13$_{-0.11}^{+0.17}$ & 1.02$_{-0.07}^{+0.09}$ & 0.97$_{-0.05}^{+0.08}$ & 1.21$_{-0.12}^{+0.13}$ \\
0.08--0.13 & 1.12$_{-0.05}^{+0.06}$ & 1.18$_{-0.08}^{+0.10}$ & 1.02$_{-0.07}^{+0.06}$ & 1.08$_{-0.08}^{+0.12}$ & 1.02$_{-0.07}^{+0.12}$ & 1.31$_{-0.13}^{+0.17}$ \\
0.13--0.20 & 1.21$_{-0.05}^{+0.06}$ & 1.21$_{-0.05}^{+0.06}$ & 1.23$_{-0.10}^{+0.08}$ & 1.19$_{-0.05}^{+0.08}$ & 1.16$_{-0.14}^{+0.10}$ & 1.25$_{-0.10}^{+0.10}$ \\
0.20--0.28 & 1.09$_{-0.05}^{+0.05}$ & 1.11$_{-0.05}^{+0.12}$ & 1.08$_{-0.06}^{+0.05}$ & 1.17$_{-0.06}^{+0.07}$ & 1.28$_{-0.07}^{+0.09}$ & 1.08$_{-0.07}^{+0.08}$ \\
0.28--0.36 & 1.12$_{-0.06}^{+0.10}$ & 1.12$_{-0.09}^{+0.19}$ & 1.10$_{-0.05}^{+0.08}$ & 1.05$_{-0.06}^{+0.07}$ & 1.01$_{-0.08}^{+0.07}$ & 1.15$_{-0.10}^{+0.09}$ \\
0.36--0.46 & 0.91$_{-0.04}^{+0.05}$ & 1.03$_{-0.10}^{+0.12}$ & 0.87$_{-0.05}^{+0.05}$ & 0.97$_{-0.08}^{+0.04}$ & 0.90$_{-0.08}^{+0.08}$ & 0.98$_{-0.08}^{+0.09}$ \\
0.46--0.58 & 0.94$_{-0.04}^{+0.08}$ & 0.89$_{-0.07}^{+0.05}$ & 1.04$_{-0.11}^{+0.12}$ & 0.83$_{-0.03}^{+0.03}$ & 0.84$_{-0.05}^{+0.08}$ & 0.88$_{-0.03}^{+0.04}$ \\
0.58--0.74 & 0.81$_{-0.04}^{+0.09}$ & 0.69$_{-0.07}^{+0.10}$ & 0.93$_{-0.04}^{+0.08}$ & 0.80$_{-0.08}^{+0.07}$ & 0.85$_{-0.15}^{+0.14}$ & 0.81$_{-0.10}^{+0.13}$ \\
0.74--0.95 & 0.67$_{-0.06}^{+0.07}$ & 0.82$_{-0.10}^{+0.12}$ & 0.57$_{-0.06}^{+0.08}$ & 0.54$_{-0.03}^{+0.04}$ & 0.52$_{-0.08}^{+0.07}$ & 0.55$_{-0.05}^{+0.07}$ \\
0.95--1.20 & 0.64$_{-0.07}^{+0.10}$ & 0.68$_{-0.08}^{+0.09}$ & 0.60$_{-0.11}^{+0.20}$ & 0.50$_{-0.10}^{+0.12}$ & 0.41$_{-0.05}^{+0.09}$ & 0.48$_{-0.10}^{+0.14}$ \\
1.20--1.50 & 0.66$_{-0.24}^{+0.33}$ & 1.11$_{-0.26}^{+0.39}$ & 0.25$_{-0.09}^{+0.19}$ & 0.45$_{-0.14}^{+0.07}$ & 0.45$_{-0.16}^{+0.16}$ & 0.42$_{-0.18}^{+0.10}$ 
\enddata
\tablecomments{All uncertainties are 1$\sigma$. 
The procedures for producing these profiles are described in detail in \S2.1--2.2.}
\label{table:univT}
\end{deluxetable*}

This method, which has $2N+1$ free parameters \emph{per annulus} (see Table \ref{table:xspec}), makes the assumption that, in the subsample of $N$ clusters, there is a universal temperature profile of the form $kT/kT_{500}$ = $f$($r/R_{500}$). We do not make any further assumptions about the form of this function. 
We simultaneously fit spectra for all $N$ clusters, including multiple observations where available, along with off-source background spectra (at $r\gtrsim3R_{500}$) unique to each cluster. 
Goodness-of-fit was determined using the $\chi^2$ parameter, with weighting based on \cite{churazov96}, which has been shown to yield unbiased parameter estimates for spectra containing as few as $\sim$50 total counts. 
We repeat this process 100 times for each annulus in order to assess the cluster-to-cluster variation in $kT/kT_{500}$. We note that, while the metallicity (Z$_i$) is left as a free parameter, we do not have sufficiently deep data to put meaningful constraints on the shape of Z($r/R_{500}$). We leave this parameter free so that, in the few clusters with a strong Fe~\textsc{xxv} emission line, the temperature is not skewed high in order to improve the fit. This method was found to yield the smallest residuals in temperature for simulated clusters with low counts and unknown metallicity\footnote{\scriptsize https://heasarc.gsfc.nasa.gov/xanadu/xspec/manual/XSfakeit.html}.

\begin{figure*}[htb]
\centering
\includegraphics[width=0.99\textwidth]{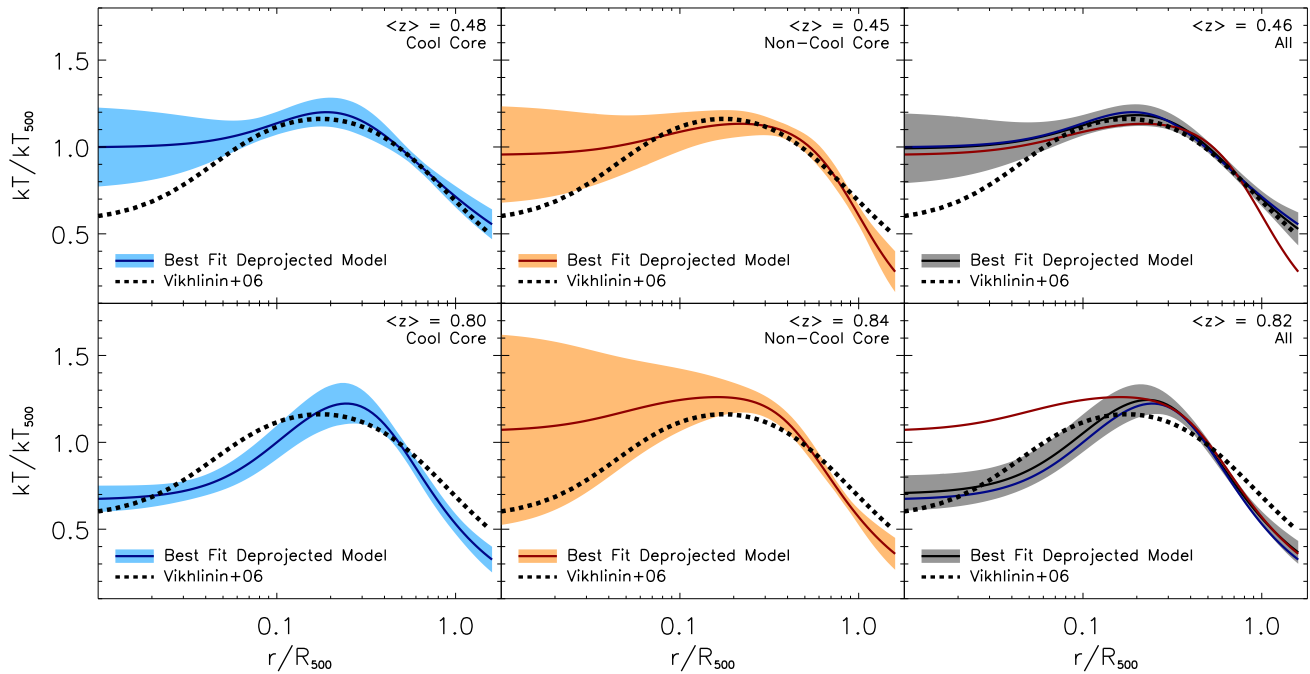}
\caption{Deprojected temperature profiles for each of the six subsamples. Colored curves correspond to the best-fit deprojected model and the 1$\sigma$ uncertainty in this model. In the right-most panels, we overplot the cool core and non-cool core profiles for comparison. In all panels we show the average deprojected temperature profile from \cite{vikhlinin06a} for comparison. This figure highlights the lower-temperature cores and outskirts in high-$z$ clusters, relative to their low-$z$ counterparts. The best-fit parameters which describe these curves are provided in Table \ref{table:ktfits}.}
\label{fig:univT_model}
\end{figure*}

In Figure \ref{fig:univT} and Table \ref{table:univT} we show the result of this joint-fit analysis for each of the 6 subsamples described in \S2.1. This figure demonstrates our ability to constrain the projected temperature to within $\sim$10\% over 100 realizations, despite the fact that the average individual cluster contributes only $\sim$100--200 X-ray counts  to each of these bins (Figure \ref{fig:counts}). The uncertainty range shown in Figure \ref{fig:univT} represents the cluster-to-cluster scatter in the $kT/kT_{500}$ profile, which dominates over the statistical uncertainty in the joint fit.
Overall the general shapes are as expected -- centrally-concentrated clusters have cool cores, while their counterparts do not. The agreement with previous average projected temperature profiles \citep[e.g.,][]{pratt07,baldi12} is qualitatively good. In particular, when the results from \cite{baldi12} are scaled to the same normalization ($r/R_{500}$, $kT/kT_{500}$), they agree at the 1$\sigma$ level with our low-$z$ subsample. 
In general, the ICM in the outskirts of high-$z$ clusters appears to have a steeper temperature gradient than the ICM in their low-$z$ counterparts.
We note, however, that the temperature profiles shown in Figure \ref{fig:univT} are \emph{projected} -- in order to determine the true radial distribution of ICM temperature, entropy, and pressure, we must first perform a three-dimensional deprojection.

\subsection{Deprojecting Mean Temperature Profiles}
To determine the three-dimensional average temperature profile, we project an analytic function onto two dimensions and fit this projected model to the data. Our three-dimensional temperature model is inspired by Eq.\ 6 from \cite{vikhlinin06a}:
\begin{equation}
\frac{k\rm{T}}{k\rm{T}_{500}} = \textrm{T}_0\frac{((x/r_c)^{a_{cool}}+(\textrm{T}_{min}/\textrm{T}_0))}{(1+(x/r_c)^{a_{cool}})}\frac{(x/r_t)^{-a}}{(1+(x/r_t)^b)^{c/b}}
\end{equation}
where $x=r/R_{500}$. This equation models the temperature profile in two parts: 1) the core region, which has a temperature decline parametrized by a minimum temperature ($T_{min}$), scale radius ($r_c$), and shape ($a_{cool}$) \citep{allen01} and 2) a broken power law with a characteristic inner slope ($a$), transition steepness ($b$), and outer slope ($c$), and a transition radius ($r_t$).
Since we only sample the cool core ($r\lesssim0.1R_{500}$) with 2--3 data points (Figure \ref{fig:univT}), we have removed two degrees of freedom from the more general parametrization presented in \cite{vikhlinin06a}, fixing $a_{cool}=2$ and $a=0$, as otherwise the fit would be completely unconstrained in the inner parts. To project this function onto two dimensions, we follow the procedures described in detail by \cite{vikhlinin06b}, which require two additional ingredients, aside from the analytic temperature profile: the three-dimensional electron density and metallicity profiles. For the latter, we assume the average metallicity profile from \cite{leccardi08b}, but confirm that radically different metallicity profiles (i.e., flat, inverted) result in $\lesssim$5\% differences in the deprojected temperature, and then only at low temperatures ($\lesssim2$ keV). For the electron density profile, we utilize the deprojected density profiles for each cluster from \cite{mcdonald13b}. Since each bootstrapped temperature profile is actually the weighted average of $N$ clusters, we compute the appropriate three-dimensional gas density profile as follows:
\begin{equation}
\left<\,\frac{\rho_g(r)}{\rho_{crit}}\,\right> = \frac{\sum\limits_{i=1}^N C_i (r) \times \frac{\rho_{g,i}(r)}{\rho_{crit}}}{\sum\limits_{i=1}^N C_i (r)} ~~,
\end{equation}
where $C_i(r)$ is the number of X-ray counts for cluster $i$ at radius $r$. This produces a mean gas density profile, weighted in approximately the same way as the mean temperature profile.

For each bootstrap realization, we use the same $N$ clusters in the calculation of both the mean temperature and density profiles. From these profiles, we project the temperature profile along a given line of sight, following \cite{vikhlinin06b}. This procedure accounts for different contributions from continuum and line emission along the line integral, providing an accurate estimate of the projected single-temperature model. To correctly factor in the temperature-sensitive detector response, we convert to absolute temperature units (from $kT/kT_{500}$) using the average $kT_{500}$ for the subsample of N clusters. For each radial bin, the projected temperature was computed by numerically integrating along the line of sight over $-4R_{500} < r < 4R_{500}$ as well as along the bin in the radial direction. Once complete, this procedure yielded a projected temperature profile which was fit to the data using a least-squares $\chi^2$ minimization routine. This process was repeated for each realization of the projected temperature profile (Figure \ref{fig:univT}), allowing an estimate of the uncertainty in the deprojected model.

\begin{deluxetable}{c c c c c c}[htb]
\tablecaption{Best-Fit Temperature Profile Parameters}
\tablehead{
\colhead{Subsample} & 
\colhead{$r_c$} & 
\colhead{T$_{min}$/T$_0$} & 
\colhead{$r_t$} & 
\colhead{$b$} & 
\colhead{$c$}
}
\startdata
low-$z$ & 0.10 &  0.77 & 0.40 & 2.79 & 0.64 \\ 
low-$z$, CC & 0.13 &  0.69 & 0.30 & 2.19 & 0.57 \\
low-$z$, NCC & 0.08 &  0.81 & 0.96 & 2.74 & 2.42 \\
\\
high-$z$ & 0.10 &  0.49 & 0.38 & 3.26 & 0.94 \\ 
high-$z$, CC & 0.11 &  0.47 & 0.41 & 3.30 & 1.08 \\ 
high-$z$, NCC & 0.05 &  0.82 & 0.46 & 3.41 & 1.02 \\ 
\enddata
\tablecomments{The functional form for these fits, which are shown in Figure \ref{fig:univT_model}, is provided in Eq.\ 2. This parametrization is adopted from \cite{vikhlinin06a}, holding $a_{cool}=2$ and $a=0$ fixed. Both $r_c$ and $r_t$ are expressed in units of $R_{500}$.}
\label{table:ktfits}
\end{deluxetable}

The resulting deprojected temperature profiles, along with the 1$\sigma$ uncertainty regions, are shown in Figure \ref{fig:univT_model} and Tables \ref{table:univT} and \ref{table:data}. These results suggest that high-redshift clusters tend to be, on average, $\sim$30--40\% cooler at large radii ($r\sim 1.5\textrm{R}_{500}$) than their low-redshift counterparts. However, when non-cool cores are considered on their own, this trend is reversed (albeit at low significance). This may be a result of, on average, more significant ``clumping'' in the outskirts of high-$z$ clusters \citep{nagai11}. Infalling subhalos will tend to have lower temperature and higher density than the surrounding ICM, leading to a bias towards low temperatures in the emission measure-weighted spectra if they are unresolved. We will discuss this possibility further in \S4, in the context of the average entropy profile.
Similar to previous works, we find that the peak temperature for relaxed (cool core) systems is reached at $\sim$0.2--0.3$R_{500}$.
The cores of high-z cool core clusters appear to be cooler than their low-z counterparts at the $\sim$30\% level, an effect still pronounced in the the full ensemble.
Given that our typical 1-$\sigma$ uncertainty in the central bin is $\sim$15\%, this is only marginally significant.

\section{Mean Pressure and Entropy Profiles}

\begin{figure*}[htb]
\centering
\includegraphics[width=0.98\textwidth]{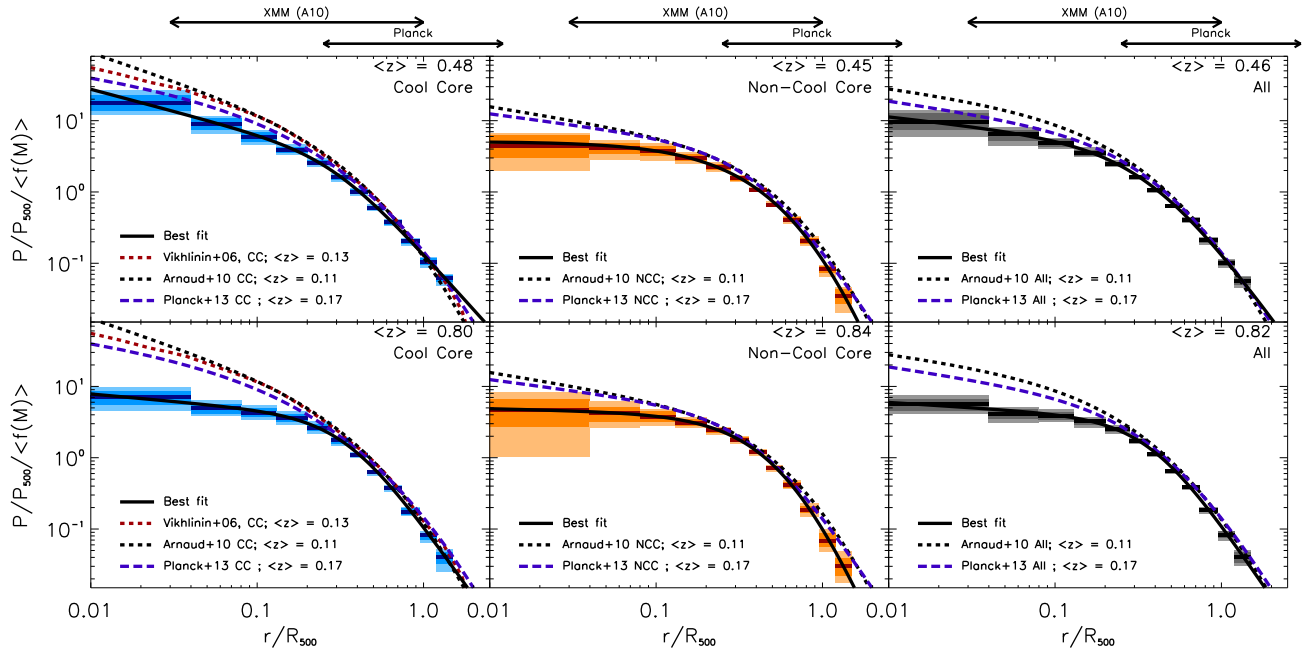}
\vskip -0.1 in
\caption{Average pressure profiles for 80 SPT-selected clusters. In all panels, the colored regions represent the data, with dark and light regions corresponding to 1$\sigma$ and 90\% confidence, respectively, while the solid horizontal line corresponds to the median value. The best fit GNFW profile (Eq.\ 8) is shown in solid black, while comparisons to V06, A10, and P13 are shown in dotted and dashed lines. The relative spatial coverage of A10 and P13 are shown at the top, for reference. For the low-$z$ subsample, the agreement between all three samples is excellent. For high-$z$ clusters, there appears to be slightly less pressure both in the core ($r<0.1R_{500}$) and outskirts ($r>R_{500}$).}
\label{fig:univP}
\end{figure*}

Previous studies (see \S1) have measured the ``Universal'' pressure and/or entropy profiles by simply taking the mean or median of a large number of individual profiles, which are individually calculated as follows: 

\begin{equation}
P(r) = n_e(r) \times kT(r), ~~~K(r) = n_e(r)^{-2/3} \times kT(r) .
\end{equation}

\noindent{}This approach is unfeasible for this sample, given that each individual cluster only has $\sim$2000 X-ray counts and, thus, the individual $kT(r)$ profiles are unconstrained. However, if we define $P_{500}$ and $K_{500}$ as follows \citep[from][]{nagai07}:

\begin{equation}
P_{500} = n_{g,500} \times kT_{500}, ~~~K_{500} = \frac{kT_{500}}{n_{e,500}^{2/3}}
\end{equation}
where $n_{g,500} = (\mu_e/\mu)n_{e,500} = 500f_b\rho_{crit}/(\mu m_p)$, and we assume that $\mu = 0.59$ is the mean molecular weight and $f_b=\Omega_b/\Omega_M \sim 0.14$ \citep{gonzalez13} is the universal baryon fraction, then we can express the scaled pressure and entropy profiles as follows:

\begin{equation}
\frac{P}{P_{500}} =  0.0073\left(\frac{kT}{kT_{500}}\right)\left(\frac{\rho}{\rho_{crit}}\right) ,
\end{equation}

\begin{equation}
\frac{K}{K_{500}} = 17.2\left(\frac{kT}{kT_{500}}\right)\left(\frac{\rho}{\rho_{crit}}\right)^{-2/3} .
\end{equation}

\noindent{}These expressions allow us to measure $P/P_{500}$ and $K/K_{500}$ as a function of radius, without actually measuring $P(r)$ or $K(r)$ for any given cluster, as is usually done. Instead, we simply combine the average $\rho(r)/\rho_{crit}$ profiles \citep{mcdonald13b} with the average $kT(r)/kT_{500}$ profile from our joint-fit analysis (see \S2). The average pressure and entropy profiles, computed using Eqs.\ 6 and 7, are provided in Table \ref{table:data}.

\subsection{Pressure}
In Figure \ref{fig:univP} we show the average pressure profile, $P(r)/P_{500}$, from this work. 
We compare our joint-fit results to previous studies by \cite{vikhlinin06a} (V06; $\left<z\right> = 0.13$), \cite{arnaud10} (A10; $\left<z\right> = 0.11$), \cite{planck13} (P13; $\left<z\right> = 0.17$), after normalizing all profiles by $f(M) = (M_{500}/3\times10^{14}h_{70}^{-1}M_{\odot})^{0.12}$, following \cite{sun11} and \cite{planck13}, in order to account for small differences in mass range between samples.
We apply an additional scaling to profiles derived based on \emph{XMM-Newton} data, which accounts for a 16\% difference in temperature normalization between \emph{Chandra} and \emph{XMM-Newton} for massive clusters \citep{schellenberger14}. Both A10 and P13 define $P_{500} \propto M_{500}^{2/3} \propto kT_{500}^{0.37}$, so a 16\% normalization error in temperature corresponds to a 10\% normalization error in $P/P_{500}$. Likewise, since R$_{500} \propto M_{500}^{1/3} \propto kT_{500}^{0.19}$, we apply a 3\% renormalization in R$_{500}$ to account for differences between the two X-ray telescopes. 


Figure \ref{fig:univP} demonstrates our ability to reach larger radii than \cite{arnaud10}, who used data from \emph{XMM-Newton}, due to the fact that the \emph{Chandra} ACIS-I field of view represents a larger physical size at high redshift. While we are unable to probe as deep into the outskirts as \emph{Planck}, the unmatched angular resolution of \emph{Chandra} allows us to sample a factor of $\sim$10 finer than the native \emph{Planck} resolution in the cluster core. Thus, this work bridges the gap between A10, who primarily samples the inner pressure profile, and P13, who primarily samples the cluster outskirts.

For our low-$z$ subsample, there is good overall agreement between our results and previous work. Specifically, at $0.2R_{500} < r < 1.5R_{500}$, our results are fully consistent at the $\sim1\sigma$ level with both A10 and P13. 
At $r<0.2R_{500}$, the cool cores from A10 appear to have higher pressure. This may be due to their different cool core classification, based on central density, which also varies as a function of total mass \citep{vikhlinin06a}. Alternatively, the difference could be due to a difference in centering -- we use the large-area centroid, while A10 use the X-ray peak -- but these should both agree for relaxed, centrally-concentrated systems. The difference in the core pressure is less dramatic when the full samples are considered. 

\begin{figure}[h!]
\centering
\includegraphics[width=0.35\textwidth, trim=-1.2cm 0cm 0.5cm 0cm]{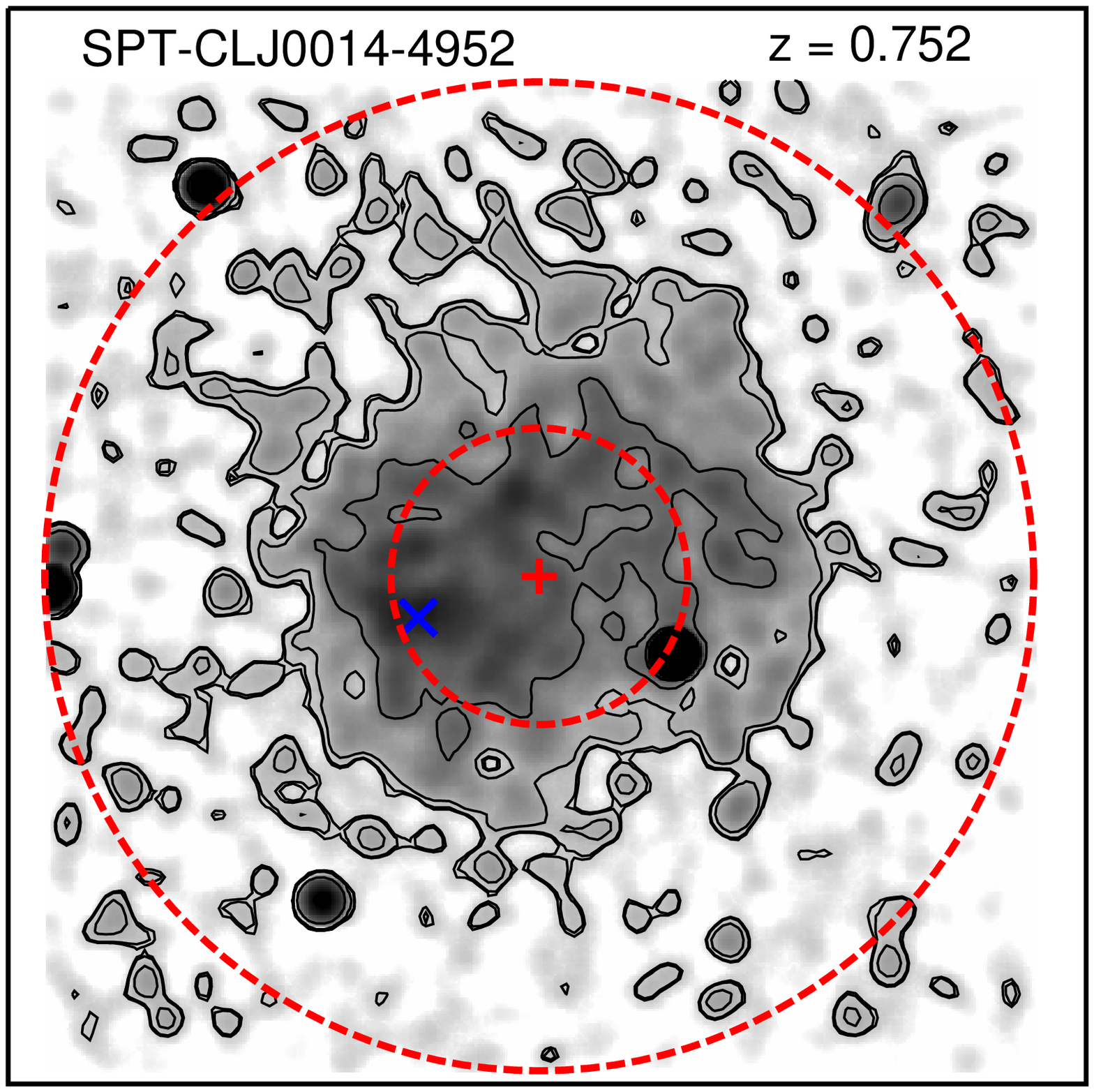}\\
\includegraphics[width=0.375\textwidth, trim=0cm 7.1cm 0cm 0cm]{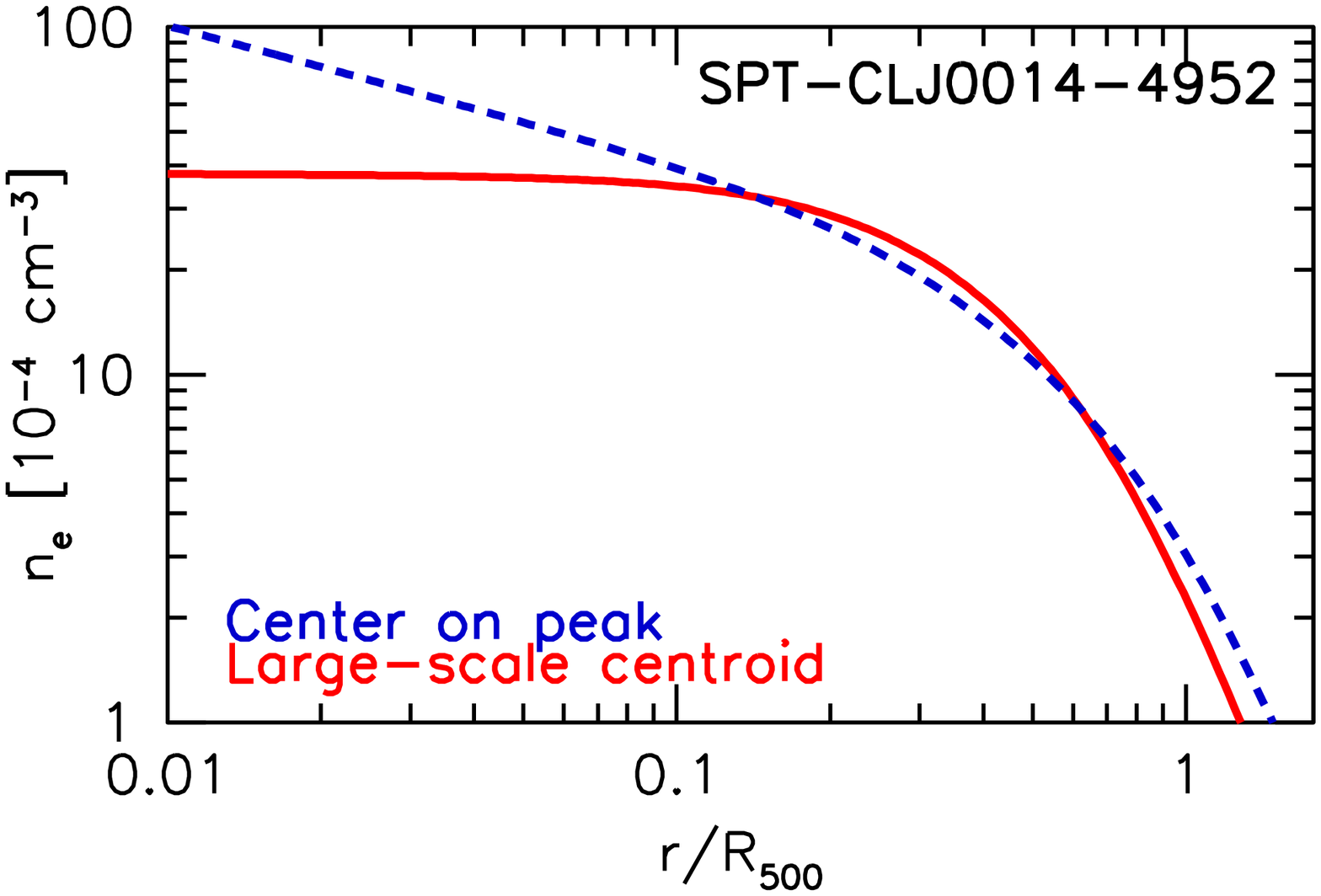}
\caption{Top: Smoothed X-ray image of SPT-CLJ0014-4952. The X-ray peak and large-annulus centroid are depicted with blue and red crosses, respectively. The two red dashed circles correspond to radii of 0.3$R_{500}$ and $R_{500}$. 
Bottom: Electron density profile, derived using the two different centers. For ``disturbed'' clusters the X-ray peak can be significantly offset from the large-scale centroid, leading to a deficit of core pressure ($\sim$50\%) and slight excess pressure at large radii ($\sim$15\%) if the latter is chosen as the cluster center.}
\vskip -0.1 in
\label{fig:centroid}
\end{figure}

\begin{deluxetable*}{c c c c c c c}[htb]
\centering
\tablecaption{GNFW Fit to Average Pressure Profiles: $P/P_{500} = P_0 (c_{500}x)^{-\gamma}[1+(c_{500}x)^{\alpha}]^{-(\beta-\gamma)/\alpha} \times \left<f(M)\right>$}
\tablewidth{500pt}
\tablehead{
\colhead{Subsample} & 
\colhead{$P_0$} & 
\colhead{$c_{500}$} & 
\colhead{$\gamma$} & 
\colhead{$\alpha$} & 
\colhead{$\beta$} &
\colhead{$\left<f(M)\right>$}
}
\startdata
\bf low-$z$ & $\bm{   4.33_{-  1.66}^{+  3.90}$} & $\bm{   2.59_{-  0.79}^{+  0.74}$} & $\bm{   0.26_{-  0.26}^{+  0.22}$} & $\bm{   1.63_{-  0.41}^{+  1.01}$} & $\bm{   3.30_{-  0.57}^{+  0.86}$} & \bf 1.070\\
A10 & 8.40 & 1.18 & 0.31 & 1.05 & 5.49$^{\dagger}$\\
P13 & 6.41 & 1.81 & 0.31$^{\dagger}$ & 1.33 & 4.13\\\\
\bf low-$z$, CC & $\bm{   3.39_{-  0.88}^{+  4.58}$} & $\bm{   3.42_{-  0.74}^{+  0.78}$} & $\bm{   0.62_{-  0.29}^{+  0.13}$} & $\bm{   2.31_{-  1.04}^{+  3.58}$} & $\bm{   2.61_{-  0.29}^{+  0.72}$} & \bf 1.064\\
A10, CC &   3.25 & 1.13 & 0.77 & 1.22 & 5.49$^{\dagger}$\\
P13, CC & 11.8 & 0.60 & 0.31$^{\dagger}$ & 0.76 & 6.58 \\\\
\bf low-$z$, NCC & $\bm{   5.10_{-  2.15}^{+  0.02}$} & $\bm{   0.88_{-  0.05}^{+  1.08}$} & $\bm{   0.00_{-  0.00}^{+  0.17}$} & $\bm{   1.23_{- 0.01}^{+  0.51}$} & $\bm{   7.58_{-  3.16}^{+  0.42}$} & \bf 1.076\\
A10, NCC & 3.20 & 1.08 & 0.38 & 1.41 & 5.49$^{\dagger}$\\
P13, NCC & 4.72 & 2.19 & 0.31$^{\dagger}$ & 1.82 & 3.62\\
\\
\bf high-$z$          & $\bm{   3.47_{-  0.67}^{+  1.09}$} & $\bm{   2.59_{-  0.38}^{+  0.37}$} & $\bm{   0.15_{-  0.15}^{+  0.13}$} & $\bm{   2.27_{-  0.40}^{+  0.89}$} & $\bm{   3.48_{-  0.39}^{+  0.60}$} & \bf 1.034\\
\bf high-$z$, CC   & $\bm{   3.70_{-  0.86}^{+  2.17}$} & $\bm{   2.80_{-  0.77}^{+  0.47}$} & $\bm{   0.21_{-  0.21}^{+  0.13}$} & $\bm{   2.30_{-  0.67}^{+  0.68}$} & $\bm{   3.34_{-  0.42}^{+  1.01}$} & \bf 1.026\\
\bf high-$z$, NCC & $\bm{   3.91_{-  1.51}^{+  0.63}$} & $\bm{   1.50_{-  0.40}^{+  0.74}$} & $\bm{   0.05_{-  0.05}^{+  0.22}$} & $\bm{   1.70_{-  0.17}^{+  0.99}$} & $\bm{   5.74_{-  1.72}^{+  2.26}$} & \bf 1.043\\
\enddata
\tablecomments{Generalized NFW fit to our average pressure profile (see Figure \ref{fig:univP}). For comparison, fits from \cite{arnaud10} and \cite{planck13} are shown. For clarity, rows that are bolded are original to this work. Values with a $^{\dagger}$ mark have been held fixed during the fitting procedure. Quoted uncertainties represent the 1$\sigma$ confidence intervals for each parameter. We note that there are strong covariances between parameters which are not encapsulated in these uncertainty ranges. For example, if all other parameters were held fixed, the uncertainty on $P_0$ would be of order $\sim$5\% for all fits.}
\label{table:gnfw}
\end{deluxetable*}

Non-cool core clusters (both high- and low-$z$) appear to have a deficit of pressure in their cores relative to A10 and P13, both at the 1--2$\sigma$ level. 
Again, this may be a result of different centering choices: A10 and, presumably, P13 choose the X-ray peak as the cluster center, while we have measured the cluster center at larger radii in order to minimize the effects of core sloshing \citep[see e.g.,][]{zuhone10}. Our choice of centering allows for the X-ray peak to be offset from the chosen center, which would lead to lower central pressure. An example of the difference between the X-ray peak and the ``large-radius centroid'' is shown in Figure \ref{fig:centroid} for SPT-CLJ0014-4952. At large radii, the cluster is clearly circular on the sky. However, the X-ray peak, presumably representing a sloshing cool core, is located at a distance of $\sim0.3R_{500}$ from the large-radius centroid. It is most likely these off-center emission peaks that are leading to the small differences between our stacked pressure profile and those of A10 and P13, specifically at $r \lesssim 0.1$R$_{500}$ and $r \gtrsim R_{500}$. We would argue that, for non-cool core clusters, our choice of the cluster center is more representative of the true dark matter distribution.

\begin{figure*}[htb]
\centering
\includegraphics[width=0.99\textwidth]{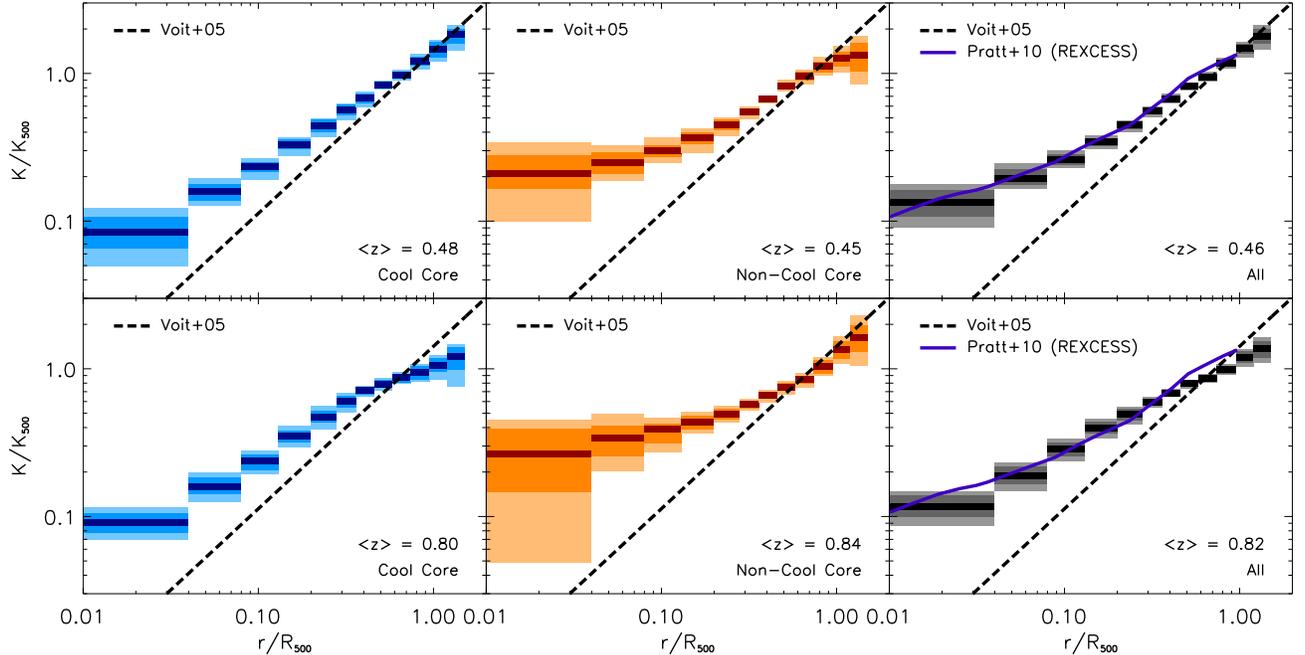}
\caption{Average entropy profiles for 80 SPT-selected clusters. In all panels, the colored regions represent the data, with dark and light regions corresponding to 1$\sigma$ and 90\% confidence, respectively, while the solid horizontal line corresponds to the median value. In all panels we show the baseline gravity-only entropy profile from simulations \citep{voit05} for comparison. In the right-most panels, we show results from \cite{pratt10} for comparison. In the low-$z$ subsample, the agreement with \cite{pratt10} is excellent, suggesting very little evolution from $z\sim0.1$ to $z\sim0.5$. At high-$z$, the entropy flattens out at large radii ($r\gtrsim R_{500}$) for the cool core subsample, as well as CC+NCC sample, which may be indicative of increased clumping in these more-distant systems. In general, the entropy profile appears to be unevolving, or evolving very slowly, at $r\lesssim 0.7R_{500}$, suggesting a remarkably stable balance between cooling and feedback.
}
\label{fig:univK}
\end{figure*}


We find that high-$z$ cool core clusters have lower central pressure than those at intermediate- and low-$z$, by factors of $\sim$3 and $\sim$6, respectively, consistent with our earlier work which showed a rapidly-evolving central gas density between $z\sim0$ and $z\sim1$, with the central, low-entropy core being less massive at $z\gtrsim0.6$ \citep{mcdonald13b}. This drop in central pressure with increasing redshift is a result of both lower central temperature (see Figure \ref{fig:univT_model}) and lower central density \citep{mcdonald13b} in relaxed, high-$z$ clusters. We note that, under the assumption of hydrostatic equilibrium, increasing the gas mass of the core ought to decrease the central temperature (assuming dark matter dominates the mass budget). The fact that we observe the opposite implies that some additional form of heating, either gravitational (e.g., increased dark matter density, adiabatic compression) or feedback-related, is raising the central temperature of low-$z$ clusters.
The observed change in central pressure would seem to suggest that high-$z$ X-ray cavities, inflated by radio-mode AGN feedback, should be a factor of $3^{1/3}$ larger in radius in order to maintain the same energetics (PdV) as low-$z$ cavities. This factor of $\sim$40\% in size evolution is currently smaller than the typical scatter in cavity size for high-$z$ clusters \citep[][Hlavacek-Larrondo \etal in prep]{hlavacek12}.

To allow a more direct comparison to A10 and P13, we fit the stacked pressure profiles with a generalized NFW (GNFW) profile following \cite{nagai07}:

\begin{equation}
\frac{P}{P_{500}}\frac{1}{f(M)} = \frac{P_0}{(c_{500}x)^{\gamma}[1+(c_{500}x)^{\alpha}]^{(\beta-\gamma)/\alpha}}
\end{equation}

\noindent{}where $x=r/R_{500}$ and $f(M)=(M_{500}/3\times10^{14} h_{70}^{-1} M_{\odot})^{0.12}$. This generalized version of the NFW profile \citep{nfw} is a broken power law with a characteristic inner slope ($\gamma$), outer slope ($\beta$), curvature ($\alpha$), and scale radius ($c_{500}$). 
The results of this fit, which is constrained over the radial range $0.01R_{500} \lesssim r \lesssim 1.5R_{500}$, are provided in Table \ref{table:gnfw}. At low-$z$, these fits are consistent with earlier work by A10 and P13. At high-$z$, these represent the first constraints on the shape of the average pressure profile, allowing a comparison to simulations over 8 Gyr of cosmic time (see \S4). 


\subsection{Entropy}
In Figure \ref{fig:univK} we show the evolution of the stacked entropy profile. Here, we compare our data to the non-radiative simulations of \cite{voit05}, rescaled to $\Delta=500$ by \cite{pratt10}. This represents the ``base'' entropy due to purely gravitational processes. The solid purple line corresponds to the median entropy profile for nearby REXCESS clusters \citep[$0.05 < z < 0.18$;][]{pratt10}, rescaled to include the cross-calibration differences between \emph{XMM-Newton} and \emph{Chandra} \citep[see \S3.1;][]{schellenberger14}. We are unable to compare to \cite{cavagnolo09} due to the lack of a normalized ($r/R_{500}$, $K/K_{500}$) median profile, or the appropriate quantities to do such a scaling ourselves.

The agreement between our low-$z$ average entropy profile and that of \cite{pratt10} is excellent, suggesting that there has been little evolution in the entropy structure of clusters since $z\sim0.6$. At large radii ($r>R_{500}$), the low-$z$ average entropy profiles are consistent with the gravity-only simulations \citep{voit05}, with the exception of the very last NCC data point, suggesting that the process responsible for injecting excess entropy at $r<R_{500}$ is unimportant at larger radii. On average, the NCC clusters have $\sim2.5\times$ higher entropy in the central bin, corresponding to a factor of $\sim$4 increase in central cooling time \citep{cavagnolo09}. 
The average CC central entropy of $\sim$0.085K$_{500}$ corresponds to $K\sim85$~keV~cm$^2$ (assuming $\left<kT_{500}\right> = 6.5$ and $\left<z\right>=0.45$), or a cooling time of $\sim3\times10^9$ yrs \citep{cavagnolo09}. We note that our CC/NCC separation is such that the CC class will contain both ``strong'' and ``weak'' cool cores, leading to a higher average cooling time \citep[see e.g.,][]{hudson10}. 

\begin{figure}[h!]
\centering
\includegraphics[width=0.49\textwidth]{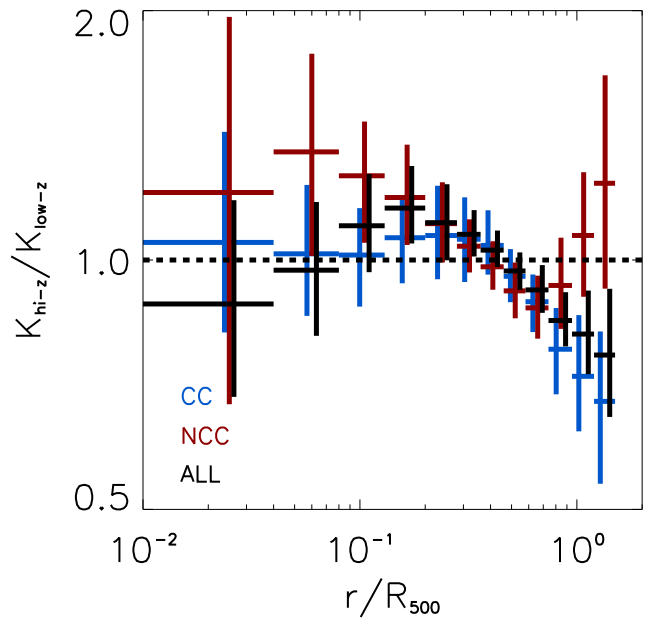}
\caption{This plot compares the low- and high-$z$ entropy profiles for three different subsamples. Here, the y-axis is the ratio of the measured entropy for the high-$z$ subsample to the low-$z$ subsample at a given radius. Error bars represent 1$\sigma$ uncertainty. All three profiles are consistent with negligible evolution at $r<0.7R_{500}$ and significant evolution at $r>0.7R_{500}$.}
\label{fig:Kevol}
\end{figure}

Consistent with \cite{mcdonald13b}, we find no significant evolution in the core entropy of CC clusters over the full redshift range studied here, despite the fact that both the central density \citep{mcdonald13b} and temperature (Figure \ref{fig:univT_model}) have evolved significantly ($\geq2\sigma$) over the same timescale. The universality of the average CC entropy profile at $r\lesssim0.7R_{500}$, shown more clearly in Figure \ref{fig:Kevol}, may be indicative of a long-standing balance between cooling and feedback processes on large scales (see also Hlavacek-Larrondo \etal in prep).

\begin{figure*}[htb]
\centering
\includegraphics[width=0.99\textwidth]{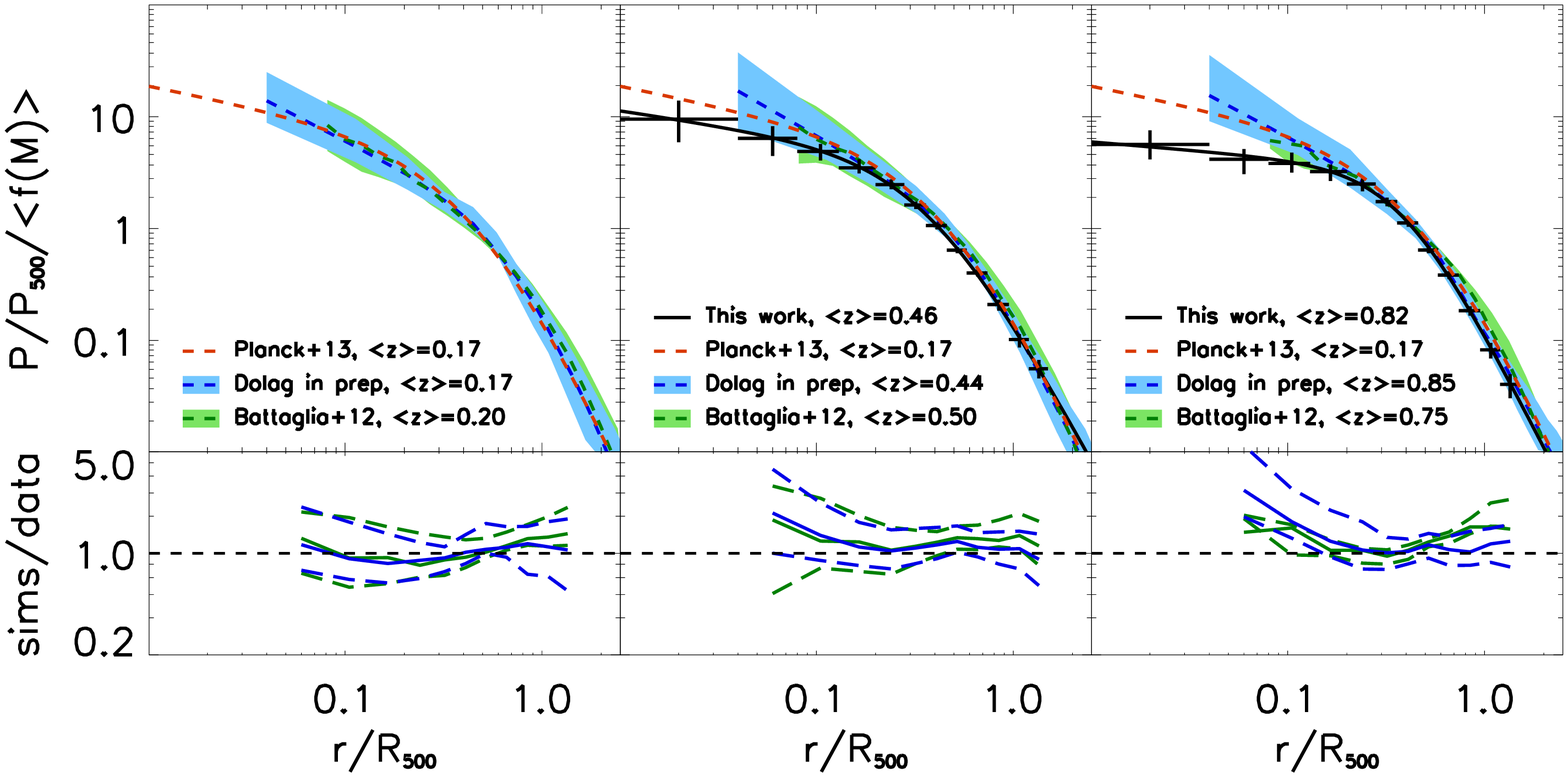}
\caption{Average pressure profile for low-$z$ ($z \lesssim 0.3$; left panels), intermediate-$z$ ($0.3<z<0.6$; middle panels), and high-$z$ ($z>0.6$; right panels) clusters. In all panels, we show the best-fit GNFW profile from \cite{planck13}, scaled down in temperature by 16\% \citep{schellenberger14} (see discussion in \S3.1). We show, for comparison, average profiles for simulated clusters with M$_{500} > 3\times10^{14}$ M$_{\odot}$ and at similar redshifts, taken from \cite{battaglia12}, and Dolag \etal (in prep). In the lower panels, we show the ratio of the simulated profiles to the data, with the 1$\sigma$ scatter depicted with dashed lines. Our best-fit GNFW profiles (see Table \ref{table:gnfw} and Figure \ref{fig:univP}) are shown with solid black lines. At all redshifts, the data agree well with the simulated clusters for $r\gtrsim0.1R_{500}$. In their cores ($r<0.1R_{500}$), simulated high-$z$ clusters from both \cite{battaglia12} and Dolag \etal (in prep) have significantly higher pressure than the data. There is mild evidence for a steepening of the pressure profile at large radii for $z>0.6$, although this may be driven by clumping in the intracluster medium near the virial radius.
}
\label{fig:univP+sims}
\end{figure*}

At $z>0.6$, we measure a distinct entropy decrement at $r>0.5R_{500}$ in high-$z$ clusters when compared to their low-$z$ counterparts. This flattening of the entropy profile is apparent in the high-$z$ CC subsample as well as the combined sample (at $>90\%$ significance in all cases), indicating that it is not being driven by one or two extreme clusters.  Figure \ref{fig:Kevol} shows that this flattening becomes significant at $r\gtrsim0.7R_{500}$.
Such a flattening could plausibly be due to clumping of the intracluster medium \citep[e.g.,][]{nagai11} -- we will discuss this possibility and the inferred clumping required to explain the measurements in the next section.

In summary, we find no significant evolution in the average entropy profile from $z\sim0.1$ \citep[][]{pratt10} to $z\sim1$ within $r\lesssim0.7R_{500}$, suggesting that the balance between cooling and feedback is exceptionally well-regulated over long periods of time ($\sim8$\,Gyr).


\section{Discussion}
Below, we discuss the results of \S2 and \S3, comparing these to previous observations and simulations in order to aid in their interpretation. In addition, we investigate potential systematic errors and/or biases which may conspire to influence our conclusions.

\subsection{Comparison to Simulations}
In Figure \ref{fig:univP+sims} we compare our average pressure profiles to those from simulations, as presented in \cite{battaglia12} and Dolag \etal (in prep). For the latter simulations, pressure profiles are computed and presented in Liu \etal (in prep). For each simulation, we show three redshift slices, similar to our low- and high-$z$ subsamples as well as a $z\sim0$ subsample for comparison to P13, and have made mass cuts similar to the SPT 2500 deg$^2$ survey selection function. All samples have been normalized by $\left<f(M)\right>$ (see \S3.1). We do not show comparisons for CC and NCC subsamples here, since i) we do not have cuspiness measurements for the simulated clusters, and ii) it is unlikely that the simulated clusters will span the full range of properties from cool core to non-cool core clusters. In general, simulations struggle to get the complicated balance between cooling and feedback right in the cores of clusters ($r<0.1R_{500}$), but perform well outside of the core where gravity dominates.

As shown in Figure \ref{fig:univP+sims}, our measured pressure profiles and the pressure profiles from both sets of simulations agree reasonably well at $r > 0.1R_{500}$ for all redshift slices. At $r<0.1R_{500}$, the difference between the simulations and the data becomes worse with increasing redshift. The simulated clusters appear to have massive cool cores in place already at $z\sim1$, while the observed clusters are becoming more centrally concentrated over the past $\sim$8\,Gyr \citep{mcdonald13b}. At large radii, the best-fit profile is consistent with Dolag \etal (in prep), and slightly steeper than that predicted by \cite{battaglia12}, but all profiles are consistent at the 1$\sigma$ level with the data. We stress that any  steepening of the pressure profile may be artificial, indicative of a bias due to clumping of the ICM at higher redshift, a point we will address below.

We can also compare our observations of an unevolving entropy core (Figure \ref{fig:univK}) to simulations, this time by \cite{gaspari12} who focus on the delicate balance between AGN feedback and cooling in the cores of simulated galaxy clusters. These simulations demonstrate that, while at the very center ($<$10\,kpc) of the cluster the entropy can fluctuate significantly (factors of $\sim$2--3) on short (Myr) timescales, the entropy at $\gtrsim$20 kpc is relatively stable of $\sim$5\,Gyr timescales. These simulations, which reproduce realistic condensation rates of cool gas from the ICM, suggest that a gentle, nearly-continuous injection of mechanical energy from the central AGN is sufficient both to offset the majority of the cooling (preventing the cooling catastrophe) and to effectively ``freeze'' the entropy profile in place.

Overall, the agreement between observations and simulations is encouraging. The primary difference between the two occurs at $r\lesssim0.1R_{500}$, with excess pressure in the simulated cores. As these are the radii where the complicated interplay between ICM cooling, bulk ICM motions, and AGN feedback is most important, it is perhaps unsurprising that the deviations between data and model are most severe in this regime.

\subsection{Cluster Outskirts: Halo Accretion?}
In recent years, a number of different studies have observed a flattening of the entropy profile for a number of different galaxy clusters at the virial radius \citep[e.g.,][]{bautz09,walker13,reiprich13,urban14}. This flattening, while not observed in all clusters \citep[e.g.,][]{eckert13}, has been attributed to clumping in the intracluster medium \citep[see e.g.,][]{simionescu11, nagai11, urban14}. If a substantial fraction of the ICM beyond the virial radius is in small, overdense clumps, the measured electron density ($n_e$) over a large annulus will be biased high, due to surface brightness being proportional to $n_e^2$. These clumps are thought to be the halos of infalling galaxies or small groups. Due to their low mass, they ought to be cooler than the ambient ICM, which could also lead to the measured temperature being biased low.

Given that we measure, on average, lower temperatures (and entropies) at large radii ($r\gtrsim R_{500}$) in high-$z$ clusters, it is worth discussing whether this result could be driven by clumping and, specifically, how massive these clumps could be. If we assume that an extended source with $<$20 X-ray counts would go undetected against the diffuse cluster emission, we can estimate a limiting X-ray luminosity at $z=0.8$ of $L_X\sim2\times10^{43}$ erg s$^{-1}$, corresponding to a halo mass of $M_{500} \sim 8\times10^{13}$ M$_{\odot}$ and temperature of $\sim$2 keV  \citep{vikhlinin09a}.  Thus, it is quite possible that the measured temperature  in the outskirts of clusters at $z>0.6$ is biased low due to our inability to detect and mask group-sized halos which are in the process of accreting onto the massive cluster. 

\begin{figure}[htb]
\centering
\includegraphics[width=0.49\textwidth]{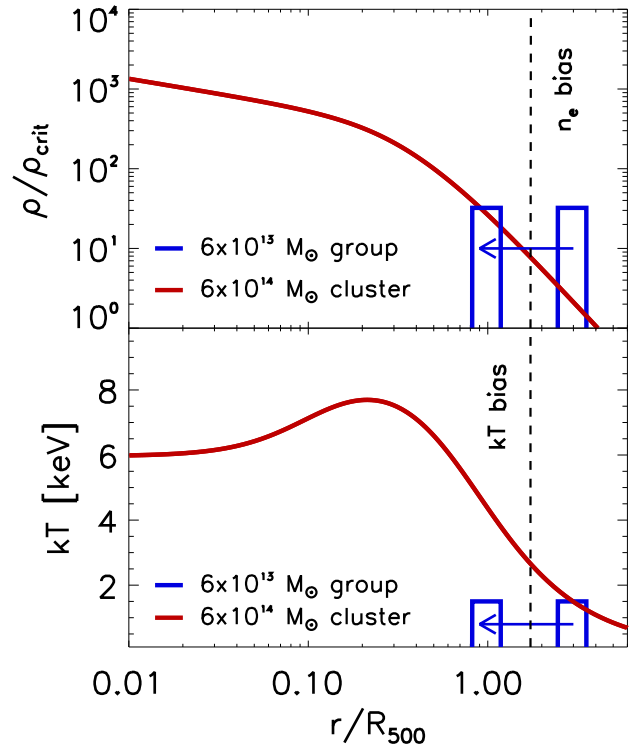}
\caption{Idealized depiction of a group-sized (M$_{500} = 6\times10^{13}$ M$_{\odot}$; blue lines) halo falling into a massive (M$_{500} = 6\times10^{14}$ M$_{\odot}$; red lines) galaxy cluster. The infalling group is assumed to be isothermal and constant density, with the density equal to $\rho_g = 0.12$M$_{500}/\frac{4}{3}\pi R_{500}^3$ and temperature taken from the M--T$_X$ relation \citep{vikhlinin09a}. This figure demonstrates that, as a group-sized halo falls into a massive cluster, it will first significantly bias the density high at $r\gtrsim 1.7R_{500}$ (right of dashed vertical line), and then bias the temperature low at $r\lesssim 1.7R_{500}$ (left of dashed vertical line). The latter effect may be driving the steep temperature profile (Figures \ref{fig:univT}--\ref{fig:univT_model}) and entropy flattening (Figure \ref{fig:univK}) that we observe in high-$z$ clusters. 
}
\label{fig:cartoon}
\end{figure}

The entropy flattening that we measure in Figure \ref{fig:univK} is driven primarily by the evolution in the temperature profile (Figures \ref{fig:univT}--\ref{fig:univT_model}), with only a small, insignificant evolution measured in the outer part of the gas density profile \citep{mcdonald13b}. This makes sense, if the temperature profile is in fact biased by infalling $>10^{13}$ M$_{\odot}$ groups at $\sim$R$_{500}$. Figure \ref{fig:cartoon} illustrates this scenario, showing the density and temperature profiles for a typical SPT-selected cluster (M$_{500} = 6\times10^{14}$ M$_{\odot}$, $kT_{500} = 6.5$ keV), and an infalling group-sized system (M$_{500} = 6\times10^{13}$ M$_{\odot}$, $kT_{500} = 1.5$ keV). For simplicity, we assume that the infalling group is isothermal and constant density, with $\rho_g = M_{g,500}/\frac{4}{3}\pi R_{500}^3$, where both M$_{g,500}$ and $R_{500}$ can be derived from the group mass, assuming a gas fraction of 0.12. This simple test shows that, at $r\gtrsim 1.7R_{500}$, group-sized halos will significantly bias the measured density high, while at $r\lesssim 1.7R_{500}$ they will bias the measured temperature low. At $\sim$R$_{500}$, where we measure a flattening of the entropy profile, the density of the infalling group and the ambient ICM are roughly equal, with a factor of $\sim$3 difference in temperature. This temperature contrast would result in an artificial steepening of the temperature profile, as we observe in Figures \ref{fig:univT}--\ref{fig:univT_model}). Following \citep{vikhlinin06b}, we estimate that the group-sized halos would need to contribute $\sim$30--40\% of the total X-ray counts in the outer annuli to bias the temperature low by the observed 40\%, with the exact fraction depending on the relative temperature of the cluster and group.

Simulations suggest that at $z\gtrsim1$ there is significantly more massive substructure in the outskirts of galaxy clusters. For example, \cite{tillson11} find that the accretion rate onto massive halos evolves by a factor of $\sim$3.5 from $z\sim1.5$ to $z\sim0$, while \cite{fakhouri10} find that  10$^{14}$ M$_{\odot}$ halos are accreting 10$^{13}$ M$_{\odot}$ subhalos at a rate $\sim$3 times higher at $z\sim1$ than at $z\sim0$. 
These results suggest that the entropy flattening which we measure (Figure \ref{fig:univK}) is consistent with a temperature bias due to our inability to detect (and mask) large substructures in the outskirts of SPT-selected clusters. We stress that this ``superclumping'' is qualitatively different than the ``clumping'' inferred in nearby clusters \citep[e.g.,][]{simionescu11,nagai11,urban14}, which is commonly interpreted as large numbers of small subhalos raining onto clusters at the virial radius, where group-sized halos would be detected and masked.

\subsection{Cool Core Evolution}
In an earlier analysis of this dataset \citep{mcdonald13b}, we saw evidence for evolution in the central gas density of cool cores over the past 8~Gyr but no evidence that the minimum ICM entropy in the central $\sim$10~kpc had evolved since $z \sim 1$, maintaining a floor at $\sim$10 keV cm$^2$. Now, with a more rigorous joint-fit analysis to constrain the central temperature, providing a more accurate estimate of the central entropy, we revisit this result.

From Figure \ref{fig:univK}, we see no measureable evolution in the central entropy bin ($0<r<0.04R_{500}$), from $K/K_{500} = 0.095_{-0.02}^{+0.02}$ at low-$z$ to $K/K_{500} = 0.102_{-0.01}^{+0.02}$ at low-$z$. Indeed, the average cool core entropy profile shows no evidence for evolution interior to $r<0.7R_{500}$ since $z\sim1$ (Figure \ref{fig:Kevol}). In the absence of feedback or redistribution of entropy, one would expect the average entropy to drop rapidly in the cores of these clusters, on Gyr or shorter timescales. Given the 5~Gyr spanned by this sample, and the consistency with the $z\sim0$ work by \cite{pratt10}, we can argue that some form of feedback is precisely offsetting cooling between $z\sim1$ and $z\sim0$. Specifically, as the central gas density increases, the core temperature also increases. This trend is contrary to what one would expect from simple hydrostatic equilibrium in a dark matter-dominated halo, but is consistent with the expectation for adiabatic compression of the gas. A likely culprit for this heat injection is radio-mode feedback \citep[e.g.,][]{churazov01, fabian12,mcnamara12}, which has been shown to be operating steadily over similar timescales \citep{hlavacek12}. Indeed, \cite{gaspari11} demonstrate that the immediate result of a burst of AGN feedback is to increase the core temperature of the gas, while leave the large-scale ($r\gtrsim0.1R_{500}$) distribution of temperatures unchanged.

We finish by stressing that this work and that of \cite{mcdonald13b} refer to the entropy in the inner $\sim$40~kpc as the ``central entropy''. This annulus, which contains all of the lowest entropy gas falling onto the central cluster galaxy, is limited in size by our relatively shallow exposures. Indeed, \cite{panagoulia13} show that with improved angular resolution the entropy continues to drop toward the central AGN. Our discussion of an ``entropy floor'' is always referring to a fixed radius, within which the mean entropy is not evolving.

\subsection{Systematic Biases/Uncertainties}
Below we briefly address three potential issues with our data analysis: whether the low signal-to-noise in cluster outskirts is driving the entropy flattening, whether joint spectral fitting yields the same results as averaging individual fits, and whether the average temperature profile is mass-dependent.

\subsubsection{X-ray Spectrum Signal-to-Noise}

While our observing program was designed to obtain 2000 X-ray counts per cluster, a variety of effects conspired to create the scatter in the observed number of counts per cluster (see Figure \ref{fig:sample}). These factors include uncertainties in the $\xi$--L$_X$ relation, uncertainties on early redshift measurements, and the presence or lack of a cool core. Here, we investigate how strongly the measured average entropy profile depends on the S/N of the included observations.

\begin{figure}[h!]
\centering
\includegraphics[width=0.49\textwidth,trim=0cm 1cm 0cm 0cm]{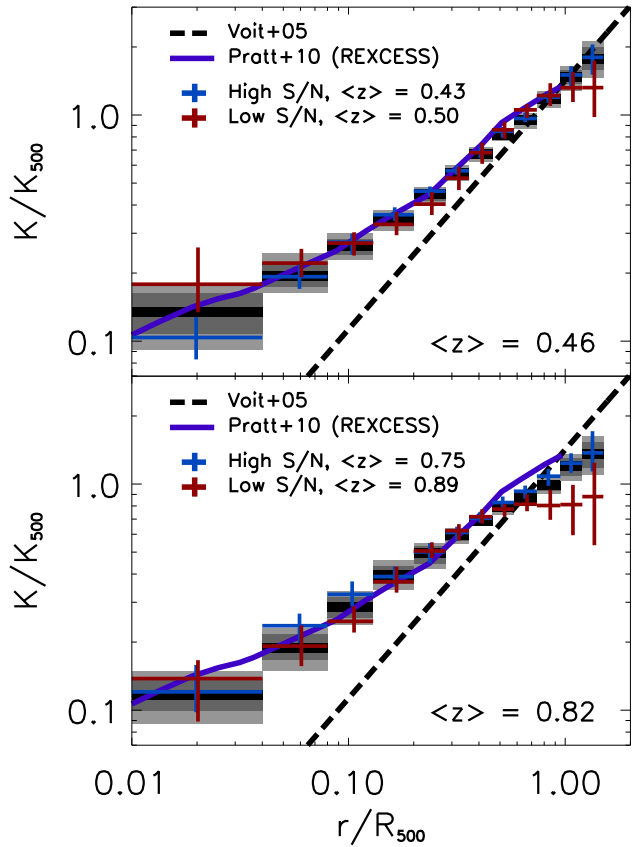}
\caption{Joint-fit entropy profile for both the low- and high-$z$ subsamples (see also Figure \ref{fig:univK}). The red and blue points correspond to the joint-fit profiles for low- and high-S/N subsamples, respectively, as described in \S4.4.1. 
We find that, at large radii, the flattening of the entropy profile correlates with both increasing redshift and decreasing S/N. The most significant flattening is present in the high-$z$, low-S/N subsample, which contains 7 of the 8 clusters at $z>1$ and all four $z>1.1$ clusters. Given that the low-$z$ and high-$z$ low-S/N subsamples have similar S/N but different degrees of flattening, we propose that the observed flattening is driven primarily by increasing redshift.
}
\label{fig:univK_bycounts}
\end{figure}

In Figure \ref{fig:univK_bycounts} we have divided the low-$z$ and high-$z$ subsamples by the S/N in the three outermost bins ($r>0.75R_{500}$), specifically so that we can test whether the observed entropy flattening is a function of S/N. For the low-S/N subsamples, there are a total of $\sim$1000 X-ray counts in each of the three outermost bins and $\sim$2700 counts per radial bin over the full radial range, compared to $\sim$2800 (outer) and $\sim$4600 (full radial range) per bin for the high-S/N subsamples. For the low-$z$ clusters, the measured entropy profile appears to be independent of the S/N -- the difference of a factor of $\sim$2 in the total number of X-ray counts used in the spectral modeling does not appear to have a significant affect on the resulting entropy profile. 
For the high-$z$ clusters, the low- and high-S/N profiles are identical at $r < 0.6 R_{500}$, with more flattening at larger radii in the low-S/N clusters.  Since the low-S/N clusters also tend to be higher redshift (the high-$z$, low-S/N subsample contains 7 of the 8 clusters at $z>1$ and all 4 clusters at $z>1.1$), it is not clear which effect is most responsible for the flattening.  In general, there is a trend of more flattening going to both higher redshift and lower S/N.  We do not expect a significant bias from low cluster counts, due to our background modeling on an observation-by-observation basis (\S 2.2), but we can not rule out this possibility.  Given that the low-$z$, low-S/N clusters have equally low S/N to the high-$z$, low-S/N clusters, we suggest that the flattening is more significantly driven by redshift evolution.

\subsubsection{Joint-Fitting Versus Profile Averaging}

\begin{figure}[htb]
\centering
\includegraphics[width=0.49\textwidth]{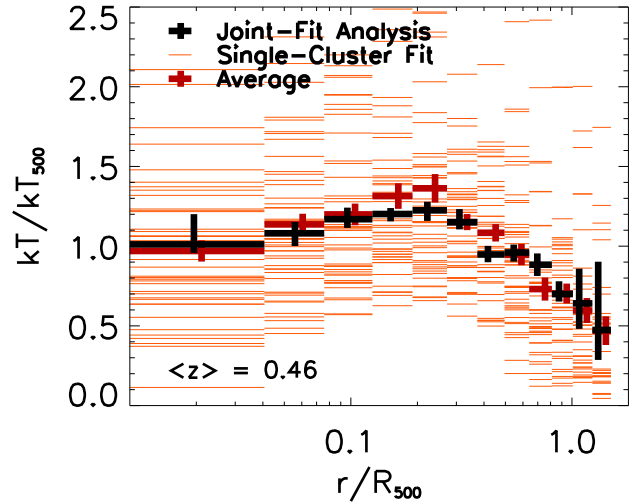}
\caption{This figure demonstrates the similarity in the average temperature profile (red) and the ``joint-fit'' profile (black; see \S2.2). Individual cluster profiles are shown as red dashes, while the average of these profiles is shown as thick red points. The uncertainty shown for the average profile is the standard error on the mean (standard deviation divided by $\sqrt{N}$) to allow a better comparison to the joint-fit uncertainties, which are measuring the scatter in the mean temperature for a large number of realizations. The joint-fit result, which is fully consistent with the average profile, is shown in black. This figure demonstrates that our joint-fit analysis is not strongly affected by combining spectra of varying signal-to-noise.}
\label{fig:allfits}
\end{figure}

To test whether our joint-fitting technique is introducing a systematic bias, we compute individual temperature profiles for our low-$z$ subsample (Figure \ref{fig:allfits}). Given that each annulus has on the order of $\sim100$ X-ray counts, these individual fits are poorly constrained. However by averaging $\sim$40 profiles (unweighted), we can constrain the average temperature profile for this subsample. For comparison, we show the results of our joint-fit analysis for the same clusters. We find that the joint-fit method and the averaging method yield consistent results. Since the uncertainty on the joint-fit analysis is really the scatter in the mean for a number of realizations (black points), we have shown the standard error on the mean (standard deviation divided by $\sqrt{N}$) in the average profile (red points) in order to make a fair comparison.

This simple test confirms that our method of joint-fitting multiple spectra is largely unbiased with respect to the true average profile. Naively, one might expect a joint-fit analysis to be biased towards the highest signal-to-noise spectra, since each cluster is essentially weighted by its total X-ray counts, while each cluster is weighted equally in the averaging method. However, this test shows that any bias that would be imparted by joint-fitting spectra of varying signal-to-noise is offset by randomly drawing and fitting subsamples of spectra.

\subsubsection{Mass Bias}
\begin{figure}[h!]
\centering
\includegraphics[width=0.49\textwidth]{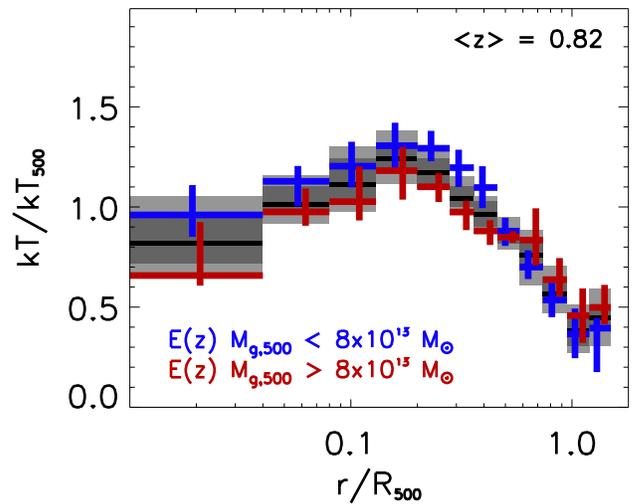}
\caption{Average temperature profiles for high-$z$ clusters. We show the combined fits in grey, low-mass systems in blue, and high-mass systems in red. This figure demonstrates that the deviation from self-similarity interior to $0.3R_{500}$, consistent with earlier work by \cite{vikhlinin06a}, is present out to $z\sim1$. Beyond $0.3R_{500}$, there is no evidence for a mass bias.}
\label{fig:univT_Mg}
\end{figure}

\cite{vikhlinin06a} show that, for a sample of relaxed, low-redshift clusters, low-mass systems tend to have higher central temperatures than their high-mass counterparts. We explore this idea in Figure \ref{fig:univT_Mg} by dividing our high-$z$ subsample by total gas mass, $M_{g,500}$, rather than by cuspiness (the following results hold for the low-$z$ subsample as well). This figure confirms that the temperature profiles of galaxy clusters are not self-similar at $r\lesssim0.3R_{500}$. We find that low mass systems have temperatures $\sim$20--30\% higher in their cores, consistent with work by \cite{vikhlinin06a} which covered a larger mass range. At $r>0.3R_{500}$ there appear to be no deviations from self-similarity, suggesting that non-gravitational processes are most likely driving the differences in the core. This figure demonstrates how important a well-selected sample is for such a joint-fit analysis to be successful and yield results representative of the true population. We expect that, given the similar mass distribution of our low- and high-$z$ subsamples (see Figure \ref{fig:sample}), this mass bias is not driving any of the trends discussed in \S3.

It is also possible that our use of the $Y_{X,500}$--M$_{500}$ relation to infer R$_{500}$ could impart a bias in these results, if the assumed evolution on this relation is incorrect. To investigate this potential bias, we re-extracted spectra using R$_{500}$ estimates based on the M$_{gas}$--M$_{500}$ relation, and repeated the analysis described in \S2.2. The resulting temperature profiles were consistent with what we have presented here, suggesting that our assumed evolution on the $Y_{X,500}$--M$_{500}$ relation is appropriate out to $z\sim1.2$.

\section{Summary}
We have presented a joint-fit analysis of X-ray spectra for 80 SPT-selected galaxy clusters spanning redshifts $0.3 < z < 1.2$. These spectra, which individually only contain $\sim$2000 X-ray counts, are divided into subsamples of $\sim$20 clusters each, and the spectra in each subsample are simultaneously modeled assuming a self-similar temperature profile. This allows us to constrain the redshift evolution of the temperature, pressure, and entropy profiles for massive clusters. Our major results are summarized below:

\begin{itemize}

\item We are able to constrain the average temperature profile out to $\sim1.5R_{500}$ for both low-$z$ ($0.2 < z<0.6$) and high-$z$ ($0.6 < z < 1.2$) clusters. The temperature profile for our low-$z$ subsample is consistent with earlier works by \cite{vikhlinin06a}, \cite{pratt07}, and \cite{baldi12}. Combined with density profiles from \cite{mcdonald13b}, we constrain the pressure and entropy profiles over $0.01R_{500} < r < 1.5R_{500}$, providing the first constraints on the redshift evolution of the Universal pressure profile.

\item The cores of high-$z$ cool core galaxy clusters appear to be marginally ($\sim$2$\sigma$) cooler than those of their low-$z$ counterparts by $\sim$30\%. This is precisely what is needed to maintain constant central entropy since $z\sim1$, given the observed evolution in the central electron density, as reported by \cite{mcdonald13b}.

\item The average temperature profile in the outskirts of high-$z$ cool core clusters is steeper than in the outskirts of low-$z$ cool core clusters. This results in a steepening of the outer pressure profile and a flattening of the outer entropy profile.
These data are consistent with an increase in the number of group-mass ($\sim$1.5 keV) halos falling into the cluster at $\gtrsim$R$_{500}$ which our relatively shallow exposures are unable to detect. This ``superclumping'' should be a factor of $\sim$3 times more common at $z\sim1$ than it is today. Failure to mask these massive subhalos can bias the temperature at $\gtrsim$R$_{500}$ low by the observed amount ($\sim$40\%).

\item The cores of low-$z$ clusters have significantly higher pressure than those of high-$z$ clusters, increasing by a factor of $\sim$10 between $z\sim1$ and $z=0$. This is driven primarily by the increase in central density with decreasing redshift \citep{mcdonald13b}, but is also boosted by the increasing central temperature with decreasing redshift.

\item We find good overall agreement between our low-$z$ average pressure profile and those of \cite{arnaud10} and \cite{planck13}.

\item Simulated clusters from \cite{battaglia12} and Dolag \etal (in prep) reproduce the evolution of the observed pressure profile at $r\gtrsim0.1R_{500}$. The growth of cool cores, resulting in a factor of $\sim$10 increase in the central pressure over the past $\sim$8\,Gyr is not reproduced in simulations.

\item We measure no significant redshift evolution in the entropy profile for cool cores at $r\lesssim0.7R_{500}$, suggesting that the average entropy profile for massive clusters is stable on long timescales and over a large range of radii. This may be a result of a long-standing balance between ICM cooling and AGN feedback.\\

\end{itemize}

\noindent{}This work demonstrates that a joint-spectral-fit X-ray analysis of low signal-to-noise cluster observations can be used to constrain the average temperature, pressure, and entropy profile to large radii. This has proven to be a powerful method for analyzing high-redshift clusters, where obtaining $>$10,000 X-ray counts per cluster is unfeasible. These techniques will add additional power to future surveys by, for example, \emph{eRosita}, or serendipitous surveys like ChaMP \citep{barkhouse06}, XCS \citep{mehrtens12}, and XXL \citep{pierre11}, where the number of clusters with data is high, but the data quality per cluster is low.

$ $

\section*{Acknowledgements} 
We thank M.\ Voit and N.\ Battaglia for helpful discussions and, along with K.\ Dolag, for sharing their simulated galaxy cluster pressure profiles.
M. M. acknowledges support by NASA through a Hubble Fellowship grant HST-HF51308.01-A awarded by the Space Telescope Science Institute, which is operated by the Association of Universities for Research in Astronomy, Inc., for NASA, under contract NAS 5-26555. 
The South Pole Telescope program is supported by the National Science Foundation through grants ANT-0638937 and PLR-1248097. Partial support is also provided by the NSF Physics Frontier Center grant PHY-0114422 to the Kavli Institute of Cosmological Physics at the University of Chicago, the Kavli Foundation, and the Gordon and Betty Moore Foundation. Support for X-ray analysis was provided by NASA through Chandra Award Numbers 12800071, 12800088, and 13800883 issued by the Chandra X-ray Observatory Center, which is operated by the Smithsonian Astrophysical Observatory for and on behalf of NASA. Galaxy cluster research at Harvard is supported by NSF grant AST-1009012 and at SAO in part by NSF grants AST-1009649 and MRI-0723073. 
The McGill group acknowledges funding from the National Sciences and Engineering Research Council of Canada, Canada Research Chairs program, and the Canadian Institute for Advanced Research.
Argonne National Laboratory's work was supported under U.S. Department of Energy contract DE-AC02-06CH11357. JHL is supported by NASA through the Einstein Fellowship Program, grant number PF2-130094.


\clearpage


\begin{appendices}
\section{Average Gas Density, Temperature, Entropy, and Pressure Profiles}
Below, we provide all of the three-dimensional deprojected quantities used in this work. Details on the derivation of these quantities, which are based on the two-dimensional temperature profiles (Table \ref{table:univT}) and the three-dimensional gas density profiles from \cite{mcdonald13b}, can be found in \S2 and \S3. All uncertainties are 1$\sigma$.
\\

\setcounter{table}{0}
\renewcommand\thetable{\Alph{section}.\arabic{table}}

\begin{table*}[h!]
\caption{Deprojected average temperature, entropy, and pressure profiles for our sample of 80 SPT-selected clusters.}
\centering
\begin{tabular}{c | c c c | c c c | c c c }
\hline\hline
 & \multicolumn{3}{c|}{All} & \multicolumn{3}{c|}{CC} & \multicolumn{3}{c}{NCC} \\
r/R$_{500}$ & $k$T$/k$T$_{500}$ & P/P$_{500}$ & K/K$_{500}$ & $k$T$/k$T$_{500}$ & P/P$_{500}$ & K/K$_{500}$ & $k$T$/k$T$_{500}$ & P/P$_{500}$ & K/K$_{500}$\\
\hline
\multicolumn{10}{c}{~}\\
\multicolumn{10}{c}{$0.3<z<0.6$}\\
0.00-0.04 & 1.02$^{+0.12}_{-0.25}$ & 10.37$^{+3.44}_{-2.44}$ &  0.13$^{+0.03}_{-0.03}$ & 1.00$^{+0.17}_{-0.19}$ & 19.28$^{+5.12}_{-4.23}$ & 0.09$^{+0.02}_{-0.02}$ & 0.99$^{+0.23}_{-0.23}$ &  4.68$^{+1.51}_{-1.33}$ & 0.22$^{+0.07}_{-0.05}$ \\
0.04-0.08 & 1.04$^{+0.09}_{-0.07}$ &  6.58$^{+0.94}_{-1.09}$ &  0.20$^{+0.02}_{-0.03}$ & 1.05$^{+0.10}_{-0.07}$ &  9.42$^{+2.00}_{-1.74}$ & 0.16$^{+0.02}_{-0.02}$ & 1.05$^{+0.11}_{-0.16}$ &  4.53$^{+0.82}_{-0.76}$ & 0.26$^{+0.03}_{-0.04}$ \\
0.08-0.13 & 1.13$^{+0.06}_{-0.08}$ &  5.17$^{+0.61}_{-0.52}$ &  0.26$^{+0.02}_{-0.02}$ & 1.14$^{+0.07}_{-0.08}$ &  6.40$^{+0.72}_{-0.86}$ & 0.23$^{+0.02}_{-0.02}$ & 1.09$^{+0.11}_{-0.09}$ &  4.04$^{+0.50}_{-0.52}$ & 0.30$^{+0.03}_{-0.03}$ \\
0.13-0.20 & 1.18$^{+0.07}_{-0.06}$ &  3.81$^{+0.28}_{-0.23}$ &  0.35$^{+0.03}_{-0.02}$ & 1.18$^{+0.08}_{-0.08}$ &  4.12$^{+0.43}_{-0.35}$ & 0.33$^{+0.03}_{-0.03}$ & 1.13$^{+0.08}_{-0.09}$ &  3.19$^{+0.33}_{-0.37}$ & 0.37$^{+0.03}_{-0.03}$ \\
0.20-0.28 & 1.17$^{+0.07}_{-0.05}$ &  2.60$^{+0.20}_{-0.15}$ &  0.44$^{+0.03}_{-0.02}$ & 1.18$^{+0.09}_{-0.09}$ &  2.69$^{+0.27}_{-0.21}$ & 0.44$^{+0.04}_{-0.04}$ & 1.11$^{+0.10}_{-0.07}$ &  2.38$^{+0.22}_{-0.21}$ & 0.44$^{+0.03}_{-0.04}$ \\
0.28-0.36 & 1.13$^{+0.06}_{-0.04}$ &  1.73$^{+0.07}_{-0.05}$ &  0.55$^{+0.03}_{-0.03}$ & 1.13$^{+0.07}_{-0.07}$ &  1.69$^{+0.15}_{-0.14}$ & 0.56$^{+0.04}_{-0.04}$ & 1.12$^{+0.04}_{-0.06}$ &  1.69$^{+0.12}_{-0.08}$ & 0.55$^{+0.02}_{-0.03}$ \\
0.36-0.46 & 1.08$^{+0.04}_{-0.03}$ &  1.14$^{+0.04}_{-0.05}$ &  0.67$^{+0.03}_{-0.03}$ & 1.07$^{+0.07}_{-0.06}$ &  1.09$^{+0.09}_{-0.08}$ & 0.68$^{+0.05}_{-0.03}$ & 1.07$^{+0.05}_{-0.04}$ &  1.14$^{+0.07}_{-0.04}$ & 0.66$^{+0.03}_{-0.03}$ \\
0.46-0.58 & 0.99$^{+0.03}_{-0.03}$ &  0.69$^{+0.03}_{-0.03}$ &  0.81$^{+0.03}_{-0.03}$ & 0.97$^{+0.04}_{-0.05}$ &  0.63$^{+0.04}_{-0.03}$ & 0.83$^{+0.06}_{-0.05}$ & 1.01$^{+0.06}_{-0.06}$ &  0.72$^{+0.04}_{-0.04}$ & 0.82$^{+0.05}_{-0.05}$ \\
0.58-0.74 & 0.90$^{+0.03}_{-0.03}$ &  0.43$^{+0.02}_{-0.02}$ &  0.95$^{+0.05}_{-0.05}$ & 0.89$^{+0.03}_{-0.04}$ &  0.40$^{+0.02}_{-0.03}$ & 0.98$^{+0.05}_{-0.05}$ & 0.92$^{+0.07}_{-0.06}$ &  0.45$^{+0.04}_{-0.03}$ & 0.96$^{+0.08}_{-0.07}$ \\
0.74-0.95 & 0.78$^{+0.04}_{-0.04}$ &  0.22$^{+0.02}_{-0.01}$ &  1.17$^{+0.07}_{-0.05}$ & 0.78$^{+0.06}_{-0.05}$ &  0.22$^{+0.02}_{-0.01}$ & 1.19$^{+0.11}_{-0.08}$ & 0.76$^{+0.05}_{-0.05}$ &  0.22$^{+0.02}_{-0.02}$ & 1.13$^{+0.09}_{-0.08}$ \\
0.95-1.20 & 0.67$^{+0.05}_{-0.05}$ &  0.11$^{+0.01}_{-0.01}$ &  1.48$^{+0.11}_{-0.14}$ & 0.70$^{+0.06}_{-0.06}$ &  0.11$^{+0.01}_{-0.01}$ & 1.52$^{+0.13}_{-0.16}$ & 0.55$^{+0.06}_{-0.06}$ &  0.08$^{+0.01}_{-0.01}$ & 1.23$^{+0.12}_{-0.12}$ \\
1.20-1.50 & 0.59$^{+0.07}_{-0.05}$ &  0.06$^{+0.01}_{-0.01}$ &  1.78$^{+0.22}_{-0.14}$ & 0.61$^{+0.08}_{-0.05}$ &  0.06$^{+0.01}_{-0.01}$ & 1.76$^{+0.22}_{-0.14}$ & 0.41$^{+0.08}_{-0.08}$ &  0.04$^{+0.01}_{-0.01}$ & 1.28$^{+0.32}_{-0.22}$ \\
\multicolumn{10}{c}{~}\\
\multicolumn{10}{c}{$0.6<z<1.2$}\\
0.00-0.04 & 0.71$^{+0.08}_{-0.08}$ &  5.69$^{+0.86}_{-0.90}$ & 0.11$^{+0.02}_{-0.02}$ & 0.68$^{+0.07}_{-0.07}$ &  7.34$^{+1.49}_{-1.65}$ & 0.09$^{+0.02}_{-0.01}$ & 1.09$^{+0.49}_{-0.56}$ &  4.58$^{+2.68}_{-2.42}$ & 0.26$^{+0.12}_{-0.13}$ \\
0.04-0.08 & 0.86$^{+0.09}_{-0.09}$ &  4.30$^{+0.53}_{-0.51}$ & 0.19$^{+0.02}_{-0.02}$ & 0.82$^{+0.07}_{-0.07}$ &  4.95$^{+1.16}_{-0.83}$ & 0.16$^{+0.02}_{-0.02}$ & 1.14$^{+0.36}_{-0.26}$ &  4.08$^{+1.45}_{-1.11}$ & 0.31$^{+0.09}_{-0.07}$ \\
0.08-0.13 & 1.04$^{+0.09}_{-0.10}$ &  3.92$^{+0.40}_{-0.57}$ & 0.28$^{+0.03}_{-0.03}$ & 1.01$^{+0.10}_{-0.09}$ &  4.56$^{+0.63}_{-0.63}$ & 0.24$^{+0.03}_{-0.03}$ & 1.26$^{+0.16}_{-0.20}$ &  3.89$^{+0.61}_{-0.64}$ & 0.38$^{+0.05}_{-0.06}$ \\
0.13-0.20 & 1.17$^{+0.14}_{-0.06}$ &  3.32$^{+0.32}_{-0.30}$ & 0.39$^{+0.04}_{-0.03}$ & 1.14$^{+0.11}_{-0.10}$ &  3.55$^{+0.53}_{-0.43}$ & 0.34$^{+0.05}_{-0.04}$ & 1.26$^{+0.12}_{-0.12}$ &  3.26$^{+0.37}_{-0.36}$ & 0.43$^{+0.04}_{-0.05}$ \\
0.20-0.28 & 1.23$^{+0.11}_{-0.07}$ &  2.57$^{+0.23}_{-0.19}$ & 0.49$^{+0.05}_{-0.03}$ & 1.22$^{+0.10}_{-0.12}$ &  2.62$^{+0.34}_{-0.17}$ & 0.46$^{+0.05}_{-0.05}$ & 1.25$^{+0.07}_{-0.08}$ &  2.55$^{+0.21}_{-0.22}$ & 0.50$^{+0.04}_{-0.04}$ \\
0.28-0.36 & 1.21$^{+0.04}_{-0.07}$ &  1.81$^{+0.10}_{-0.11}$ & 0.60$^{+0.02}_{-0.04}$ & 1.21$^{+0.12}_{-0.11}$ &  1.80$^{+0.15}_{-0.19}$ & 0.60$^{+0.05}_{-0.05}$ & 1.20$^{+0.04}_{-0.06}$ &  1.83$^{+0.14}_{-0.10}$ & 0.58$^{+0.03}_{-0.04}$ \\
0.36-0.46 & 1.11$^{+0.04}_{-0.03}$ &  1.16$^{+0.05}_{-0.04}$ & 0.69$^{+0.03}_{-0.02}$ & 1.09$^{+0.07}_{-0.06}$ &  1.11$^{+0.08}_{-0.07}$ & 0.70$^{+0.04}_{-0.04}$ & 1.13$^{+0.05}_{-0.07}$ &  1.27$^{+0.10}_{-0.07}$ & 0.67$^{+0.04}_{-0.04}$ \\
0.46-0.58 & 0.96$^{+0.04}_{-0.03}$ &  0.67$^{+0.03}_{-0.03}$ & 0.79$^{+0.03}_{-0.03}$ & 0.96$^{+0.04}_{-0.06}$ &  0.64$^{+0.04}_{-0.03}$ & 0.80$^{+0.04}_{-0.04}$ & 0.98$^{+0.08}_{-0.06}$ &  0.75$^{+0.06}_{-0.05}$ & 0.76$^{+0.06}_{-0.04}$ \\
0.58-0.74 & 0.83$^{+0.03}_{-0.03}$ &  0.40$^{+0.02}_{-0.02}$ & 0.87$^{+0.04}_{-0.03}$ & 0.82$^{+0.06}_{-0.06}$ &  0.38$^{+0.03}_{-0.02}$ & 0.86$^{+0.07}_{-0.06}$ & 0.85$^{+0.04}_{-0.05}$ &  0.44$^{+0.03}_{-0.03}$ & 0.83$^{+0.08}_{-0.05}$ \\
0.74-0.95 & 0.66$^{+0.04}_{-0.02}$ &  0.19$^{+0.01}_{-0.01}$ & 1.00$^{+0.05}_{-0.05}$ & 0.64$^{+0.05}_{-0.05}$ &  0.19$^{+0.01}_{-0.01}$ & 0.95$^{+0.06}_{-0.08}$ & 0.68$^{+0.04}_{-0.03}$ &  0.19$^{+0.03}_{-0.02}$ & 1.04$^{+0.08}_{-0.09}$ \\
0.95-1.20 & 0.53$^{+0.04}_{-0.04}$ &  0.08$^{+0.01}_{-0.01}$ & 1.14$^{+0.12}_{-0.08}$ & 0.50$^{+0.06}_{-0.07}$ &  0.08$^{+0.01}_{-0.01}$ & 1.06$^{+0.13}_{-0.13}$ & 0.55$^{+0.05}_{-0.04}$ &  0.08$^{+0.01}_{-0.02}$ & 1.35$^{+0.21}_{-0.18}$ \\
1.20-1.50 & 0.44$^{+0.06}_{-0.06}$ &  0.04$^{+0.01}_{-0.01}$ & 1.37$^{+0.22}_{-0.16}$ & 0.40$^{+0.05}_{-0.07}$ &  0.04$^{+0.01}_{-0.01}$ & 1.21$^{+0.17}_{-0.18}$ & 0.45$^{+0.07}_{-0.09}$ &  0.03$^{+0.01}_{-0.01}$ & 1.69$^{+0.52}_{-0.34}$ \\
\hline
\end{tabular}
\label{table:data}
\end{table*}

\end{appendices}
\end{document}